\documentclass[tightenlines,aps,prb,amssymb,amsfonts,eqsecnum,showpacs,floatfix,twocolumn]{revtex4}

\usepackage[T1]{fontenc}
\usepackage[latin1]{inputenc}
\usepackage{graphicx}
\usepackage{amssymb}

\makeatletter

\usepackage{bm}
\usepackage{subfigure}
\usepackage{graphicx}

\makeatother
\begin{document}

\pacs{73.23.-b,85.85.+j,72.70.+m,03.65.Yz}

\title{Statistics of charge transfer in a tunnel junction coupled to
 an oscillator}

\author{J. Wabnig}
\affiliation{Department of Physics, Ume{\aa}  University, SE-901 87 Ume\aa}
\author{D. V. Khomitsky}
\altaffiliation[Present address: ]{Nizhny Novgorod State University, 603950 Nizhny Novgorod, Russia.}
\affiliation{Department of Physics, Ume{\aa}  University, SE-901 87 Ume\aa}
\author{J. Rammer}
\affiliation{Department of Physics, Ume{\aa}  University, SE-901 87 Ume\aa}
\author{A. L.  Shelankov}
\affiliation{Department of Physics, Ume{\aa}  University, SE-901 87 Ume\aa}
\affiliation{A. F. Ioffe Physico-Technical Institute, 194021 St. Petersburg, Russia.}
\date{9 September 2005}

\begin{abstract}
The charge transfer statistics of a tunnel junction coupled to a quantum
object is studied using the charge projection technique.
The joint dynamics of
the quantum object and the number of charges transferred
through the junction is described by
the charge specific density matrix.
The method allows evaluating the joint probability distribution of the 
state of the quantum object and the
charge state of the junction.
The statistical properties of the junction current are derived
from the charge transfer statistics using the master equation for the
charge specific density matrix.
The theory is applied to a nanoelectromechanical system, 
and the influence on the average current and the current noise
of the junction is obtained
 for coupling to a harmonic oscillator.
\end{abstract}
\maketitle

\section{Introduction}

In recent years it has become possible to couple charge dynamics of
electrons to vibrational modes of a nanostructure, and the new field
of nanoelectromechanics has emerged.\cite{Ble04} Nanoelectromechanical
devices are expected to lead to new technologies, such as ultra-small
mass detection techniques,\cite{IliCraKry04,EkiHuaRou04} as well as 
stimulating fundamental studies of quantum phenomena in macroscopic
systems.\cite{MarSimPen03} For example, experiments have probed
high-frequency nanomechanical resonators in order to reach the quantum
uncertainty limit.\cite{KnoCle03,LahBuuCam04,GaiZolBad05,CleAldDri02}
Charge transfer by mechanical motion has been studied in experiments
shuttling single electrons.  \cite{SheGorJon05,ErbWeiZwe01,SchBli04}
Nanomechanical resonators are investigated for use in quantum
information processing.\cite{ArmBleSch02,GelCle05}

The simplest possibility of detection and control of the vibrational
degree of freedom in a nanomechanical system is at present via the
coupling to a quantum point contact or a tunnel junction.  Recent
theoretical studies of nanoelectromechanical systems have therefore
considered the coupling of a harmonic oscillator to a tunnel junction.
The master equation for the oscillator was originally obtained by
Mozyrsky and Martin,\cite{MozMar02} using a method limiting
considerations to the zero-temperature case, and the influence of the
coupled oscillator on the average current was obtained.  Smirnov,
Mourokh and Horing examined the non-equilibrium fluctuations of an
oscillator coupled to a biased tunnel junction.\cite{SmiMouHor03}
Clerk and Girvin derived the master equation for the oscillator, and
considered the current noise power spectrum in the shot noise regime,
studying both the {\em dc} and {\em ac} cases.\cite{CleGir04} Armour,
Blencowe and Zhang considered the dynamics of a classical oscillator
coupled to a single electron transistor,\cite{ArmBleZha04} and Armour
the induced current noise.\cite{Arm04} 
Related models were studied in molecular electronics.\cite{MitAleMil04}

Tunnel junctions functioning as position detectors or being monitored
by a vibrational mode put emphasis on a description of the current
properties of a tunnel junction in terms of its charge dynamics.
Recently we considered a general many-body system coupled to a quantum
object and considered their joint dynamics in the charge
representation.\cite{RamSheWab04} This approach, based on a previously
introduced charge projection technique,\cite{SheRam03} provides a
quantum description of charge dynamics based directly on the density
matrix for the system, and allows us to treat the number of particles
in a given piece of material as a quantum degree of freedom,
establishing thereby in proper quantum mechanical context the charge
representation.  The evolution of the coupled systems is described in
terms of the charge specific density matrix for the quantum object,
$\hat{\rho}_n(t)$, i.e., the dynamics conditioned on the number $n$ of
charges in a specified spatial region of the environment.  When a
many-body environment is coupled to another quantum object, the method
allows evaluating at any moment in time the joint probability
distribution describing the quantum state of the object and the number
of charges in the chosen region of the many-body system.  The charge
specific density matrix description of the dynamics of a quantum
object is therefore an optimal tool to study transport in
nanostructures since in electrical measurements any information beyond
the charge distribution is irrelevant.  So far we have applied the
method to charge counting in a tunnel junction coupled to a discrete
quantum degree of freedom, viz. that of a two-level system, and shown
that the charge state of the junction can function as a meter
providing a projective measurement of the quantum state of the
two-level system.\cite{RamSheWab04}

In this paper we shall apply the charge projection method to the case
where the quantum object coupled to the junction is a continuous
degree of freedom.  In particular, we shall concentrate on the
properties of the current through the junction due to the coupling to
the quantum object.  The statistical properties of the current through
the junction and its correlations with the dynamics of the quantum
object coupled to it, shall be expressed through the charge specific
density matrix.  We shall illustrate the results for the case of a
harmonic oscillator coupled to the tunnel junction.  It should be
noticed, that current experimental setups studying
nanoelectromechanical systems are operated under conditions where
temperature, oscillator excitation energy, and voltage bias across the
junction are comparable.\cite{EkiHuaRou04,KnoCle03,LahBuuCam04}
Furthermore, nanomechanical oscillators, such as a suspended beam, are
in addition to the charge dynamics of the electrons in the junction
invariably coupled to a thermal environment, say the substrate upon
which the oscillator is mounted.  We are thus considering the
situation where a quantum object in addition to interacting with a
heat bath is interacting with an environment out of
equilibrium. Having the additional parameter, the voltage,
characterizing the environment in non-equilibrium, gives rise to
features not present for an object coupled to a many-body system in
equilibrium.  The presented approach is applicable in a broad region
of temperatures and voltages of the junction and arbitrary
frequency of the oscillator and thus generalizes previous treatments.
 
The paper is organized as follows: In Sec. II we introduce the model
Hamiltonian for a generic electromechanical nanoresonator, a harmonic
oscillator coupled to a tunnel junction, and derive the Markovian
master equation for the charge specific density matrix. The master
equation for the charge unconditional density matrix, i.e., the charge
specific density matrix traced with respect to the charge degree of
freedom of the junction, is discussed in detail for the case of a
harmonic oscillator coupled to a tunnel junction.  In Sec. III we
consider the influence of the oscillator on the current-voltage
characteristic of the junction.  In Sec. IV we consider the properties
of the stationary state of the oscillator. We calculate the heating of
the oscillator due to the nonequilibrium state of the junction, and
calculate the steady state I-V characteristic of the junction.  In
Sec. V we consider the current noise in the junction using the charge
representation, and obtain the explicit expression for the
current-current correlator in the Markovian approximation.  In
Sec. VI, the theory is then applied to the case of an oscillator
influencing the current noise of the junction.  Finally, in Sec. VII
we summarize and conclude.  Details of calculations are presented in
appendices.

\section{Master equation}
\label{REDUCED MASTER EQUATION}

As a model of a nanoelectromechanical system we consider a harmonic
oscillator coupled to a tunnel junction. The transparency of the
tunnel barrier is assumed perturbed by the displacement, $x$, of the
oscillator.  The resulting Hamiltonian is
\begin{equation}
\hat{H}=\hat{H}_{0}+H_{l}+H_{r}+\hat{H}_{T}
\label{ham}
\end{equation}
where $\hat{H}_{0}$ is the Hamiltonian for the isolated harmonic
oscillator with bare frequency $\omega_{B}$ and mass $m$. A hat marks
operators acting on the oscillator degree of freedom.  The
Hamiltonians $H_{l,r}$ specify the isolated left and right electrodes
of the junction
\begin{equation}
H_{l} = 
\sum\limits
_{\mathbf{l}}\varepsilon_{\mathbf{l}}\,
c_{\mathbf{l}}^{\dagger}c_{\mathbf{l}}\quad,\quad
H_{r}=\sum\limits
_{\mathbf{r}}\varepsilon_{\mathbf{r}}\,
c_{\mathbf{r}}^{\dagger}c_{\mathbf{r}}
\label{hamlr}
\end{equation}
where $\mathbf{l,}\mathbf{r}$ labels the quantum numbers of the
single particle energy eigenstates in the left and right electrodes,
respectively, with corresponding energies
$\varepsilon_{\mathbf{l},\mathbf{r}}$.  The operator $\hat{H}_{T}$
describes the tunnelling,
\begin{equation}
\hat{H}_{T}=\hat{\mathcal{T}}+\hat{\mathcal{T}}^{\dagger}
\quad,\quad
\hat{\mathcal{T}}=\sum\limits
_{\mathbf{l},\mathbf{r}}
\hat{T}_{\mathbf{lr}}c_{\mathbf{l}}^{\dagger}c_{\mathbf{r}}
\label{hamtum1}
\end{equation}
with the tunneling amplitudes, $\hat{T}_{\mathbf{lr}} =
\hat{T}_{\mathbf{rl}}^{\dagger}$, depending on the oscillator degree of
freedom. Due to the coupling, the tunnelling amplitudes and thereby the
conductance of the tunnel junction depend on the state of the
oscillator. In the following we assume a linear coupling between the
oscillator position and the tunnel junction
\begin{equation}
\hat{T}_{\mathbf{lr}}=v_{\mathbf{lr}}+w_{\mathbf{lr}}\hat{x}
\label{tlin}
\end{equation}
where $v_{\mathbf{lr}}=v_{\mathbf{rl}}^{*}$ is the unperturbed
tunneling amplitude and $w_{\mathbf{lr}}=w_{\mathbf{rl}}^{*}$ its
derivative with respect to the position of the oscillator. The
derivation of the equation of motion for the charge specific density
matrix presented in appendix \ref{XConditional master equation}
shows that the following combinations
of the model parameters $v_{\mathbf{lr}}$ and $w_{\mathbf{lr}}$ enter
the master equation:
\begin{equation}
\left\{ \begin{array}{c}
 G_{0}\\
 G_{xx}\\
 G_{x}\\
 g_{x}\end{array}\right\} =\frac{2\pi}{\hbar}\sum\limits
 _{\mathbf{lr}}
\left\{ \begin{array}{c}
|v_{\mathbf{lr}}|^{2}\\
|w_{\mathbf{lr}}|^{2}\\
\Re\left(v_{\mathbf{lr}}^{*}w_{\mathbf{lr}}\right)\\
\Im\left(v_{\mathbf{lr}}^{*}w_{\mathbf{lr}}\right)\end{array}\right\}
\left(-\frac{\partial
 f(\varepsilon_{\mathbf{l}})}{\partial\varepsilon_{\mathbf{l}}}\right)
\delta(\varepsilon_{\mathbf{l}}-\varepsilon_{\mathbf{r}}).
\label{cond}
\end{equation}
These lumped parameters for the junction have the following physical
meaning: Let ${\sf G}(x)=e^2G(x)$, $e$ being the electron charge,
denote the conductance as a function of the oscillator coordinate $x$
when it is treated as a classical variable.  Then, $G_{0}$ gives the
conductance of the junction in the absence of coupling to the
oscillator, $G_{0}= \frac{1}{e^2}{\sf G}|_{x=0}$, and $
G_{x}=\frac{1}{2e^2}\frac{d {\sf G}}{dx}|_{x=0}$ and $
G_{xx}=\frac{1}{2e^2}\frac{d^{2} {\sf G}}{dx^{2}}|_{x=0}$.  The
coupling constant $g_{x}$ cannot be expressed via $ {\sf G}(x)$. Note
that $g_{x}$ changes its sign upon interchange of tunneling amplitudes
between the states in the two electrodes, {\it i.e.}, after the
substitution $\mathbf{l}\leftrightarrow\mathbf{r}$.  Therefore, it is
only finite for an asymmetric junction and is a measure of the
asymmetry. As shown in section \ref{Current-voltage characteristic},
$g_{x}$ generates effects similar to charge pumping, as well as
nontrivial features in the electric current noise as discussed in
section \ref{corrr}.  For later use it is convenient to present the
coupling constants in terms of {\em conductances} by introducing the
characteristic length of the oscillator
\begin{equation}
\tilde{G}_{xx}=G_{xx} x_{0}^2
\;,\;
\tilde{G}_{x}=G_{x} x_{0}
\;,\;
\tilde{g}_{x}=g_{x} x_{0}
\;,
\label{md2}
\end{equation}
where $x_{0} = (\hbar/m \omega_0)^{1/2}$, and $\omega_0$
is the frequency of the coupled oscillator as introduced in
appendix \ref{XConditional master equation}.

\subsection{Charge specific master equation}

To study the interaction of charge dynamics in a tunnel junction with
the dynamics of a quantum object, we describe the combined system, the
quantum degree of freedom coupled to a tunnel junction, using the
charge specific density matrix method introduced in
Ref. \onlinecite{RamSheWab04}.  The approach employs charge projectors
to study the dynamics of the quantum object conditioned on the charge
state of the junction. The charge projection operator,
$\mathcal{P}_{n}$, projects the state of the conduction electrons in
the junction onto its component for which exactly $n$ electrons are in
a given spatial region, say in the left electrode. The charge specific
density matrix is then specified by
\begin{equation}
\hat{\rho}_{n}(t)=\textrm{Tr}_{el}\,(\mathcal{P}_{n}\,
\rho(t))
\label{Y10sd}
\end{equation}
where $\rho(t)$ is the full density matrix for the combined system,
and $\textrm{Tr}_{el}$ denotes the trace with respect to the
conduction electrons in the junction.  Provided the system at the
initial time, $t=0$, is in a definite charge state, {\it i.e.},
described by a charge specific density matrix of the form
$\hat{\rho}_{n}(t)= \delta_{n0}\, \hat{\rho}_{0}$, where $
\hat{\rho}_{0}$ is the initial state of the quantum object, the charge
index $n$ can be interpreted as the number of charges
\emph{transferred} through the junction. Thus the charge projector
method provides a basis for charge counting statistics in the cases
where the distribution function for \emph{transferred} charge is
relevant as discussed in Ref. \onlinecite{SheRam03}.  The charge
specific density matrix allows therefore the evaluation, at any moment
in time, of the joint probability of the quantum state of the object
and the number of charges transferred through the junction. For
example, if the charge specific density matrix is traced over the
quantum object degree of freedom, the probability $p_{n}(t)$ that $n$
charges in time span $t$ are \emph{transferred} through the low
transparency tunnel junction is the expectation value of the charge
projector, or expressed in terms of the charge specific density matrix
\begin{equation}
p_{n}(t)= {\rm Tr}\,(\rho_{n}(t))
\label{jsd}
\end{equation}
where the trace is with respect to the degree of freedom of the
coupled quantum object.

The Markovian master equation for the charge specific density matrix,
$\hat{\rho}_{n}(t)$, for the case of coupling of the junction to a
quantum object is derived and discussed in appendix \ref{XConditional
master equation}. The Markovian approximation is valid for describing
slow time variations of the density matrix; the exact conditions of the
applicability are specified later once the characteristic times of the
problem have been identified.  To lowest order in the tunneling, the
master equation for the charge specific density matrix can be
generally written in terms of super-operators: a Lindblad-like term
$\Lambda$, a diffusion term $\mathcal{D}$ and a drift term
$\mathcal{J}$: \cite{RamSheWab04}
\begin{equation}
\dot{\hat{\rho}}_{n}=-\frac{i}{\hbar}[\hat{H}_{0},\hat{\rho}_{n}]
+\Lambda\{\hat{\rho}_{n}\}+\mathcal{D}\{\hat{\rho}_{n}''\}+\mathcal{J}\{\hat{\rho}_{n}'\},
\label{conmeq}
\end{equation}
where $\hat{\rho}_{n}'$ and $\hat{\rho}_{n}''$ denote the discrete
derivatives,
\begin{eqnarray}
\hat{\rho}_{n}' & = & \frac{1}{2}\left(\hat{\rho}_{n+1}-\hat{\rho}_{n-1}\right),\label{firstderiv}\\
\hat{\rho}_{n}'' & = &
\hat{\rho}_{n+1}+\hat{\rho}_{n-1}-2\hat{\rho}_{n}.
\label{deriv}
\end{eqnarray}
General expressions for the super-operators in Eq.~(\ref{conmeq}) are
presented in appendix \ref{XConditional master equation} as well as
their specific form for the case of coupling to an oscillator. The
equation shall in Sec. \ref{stationary} 
be used to study the current noise in the
junction due to the coupling to a quantum object, before we in Sec. 
\ref{noise}
consider the explicit case of an oscillator coupled to the junction.
However, first we analyze the master equation for the unconditional
density matrix, i.e., the charge specific density matrix traced with
respect to the charge degree of freedom of the junction.

\subsection{Unconditional Master equation}\label{uncond}

Often interest is not in the detailed information of the charge
evolution of the tunnel junction contained in the charge specific
density matrix. If for example interest is solely in properties of the
oscillator, this information is contained in the traced charge
specific density matrix.
We are thus led to study the master equation for the reduced or charge
unconditional density matrix, the density matrix traced with respect
to the charge degree of freedom, $\hat{\rho}(t)=\sum_{n}\hat{\rho}_{n}(t)$.
Performing the charge trace on Eq.~(\ref{conmeq}), the master equation
for the reduced density matrix can be written in the form
\begin{eqnarray}
\dot{\hat{\rho}}(t)&=&\frac{1}{i\hbar}[\hat{H}_{R},\hat{\rho}]
+\frac{\gamma}{i\hbar}\left[\hat{x},\left\{
    \hat{p},\hat{\rho}\right\}
\right]-\frac{D}{\hbar^{2}}[\hat{x},[\hat{x},\hat{\rho}]]
\nonumber
\\
&+&
\frac{A}{\hbar^{2}}\left[\hat{x},\left[\hat{p},\hat{\rho}\right]\right]
. 
\label{meqfin}
\end{eqnarray}
The form of the master equation is generic to any continuous quantum
degree of freedom coupled linearly to the junction, and has the
well-known form for a particle coupled to a heat
bath.\cite{CalLeg83,BrePet02} In the following we consider the model
Hamiltonian for a nanoelectromechanical system introduced in section
\ref{REDUCED MASTER EQUATION}, and encounter the renormalized
oscillator Hamiltonian
\begin{equation}
\hat{H}_{R}=\frac{\hat{p}^{2}}{2m}+
\frac{m\omega_{0}^{2}\hat{x}^{2}}{2}
\label{hren}
\end{equation}
which in addition to having a renormalized oscillator frequency
$\omega_{0}$, suffers a voltage dependent linear shift in the
equilibrium position of the oscillator, which in the following is
assumed absorbed into the position of the oscillator (for details see
appendix \ref{XConditional master equation}).

The second and third term on the right in Eq.~(\ref{meqfin}) represent
the physical influences of friction and fluctuations of the
environment. For a nanomechanical object, the environment consists of
several parts. The first one, which we have explicitly included in the
model, is the tunnel junction. The other one is included
phenomenologically by introducing $\gamma_{0}$ and $D_{0}$, the values
of the friction and diffusion parameters in the absence of coupling to
the junction.  The physical mechanism generating the friction
coefficient $\gamma_{0}$ and diffusion coefficient $D_{0}$ is, e.g.,
the heat exchange of the nanoscale oscillator and the bulk substrate
it is mounted on.  Thus the latter environment could also be modeled
microscopically in the standard manner of coupling a quantum object to
a heat bath.  \cite{FeyVer63,CalLeg83} 
Then, 
with the model Hamiltonian in Eq.~(\ref{ham}) giving the electronic environment
contribution due to the coupling to the junction, the total friction
and diffusion coefficients become
\begin{equation}
 \gamma = \gamma_{0} + \gamma_{e} \quad,\quad 
D = D_{0} + D_{e}
\label{uae}
\end{equation}
where $\gamma_{e}$ is the electronic contribution to the damping
coefficient
\begin{equation}
\gamma_{e}=\frac{\hbar^{2} G_{xx}}{m}
\label{gamma}
\end{equation}
 proportional to the coupling strength $ G_{xx}$, and the electronic
contribution to the diffusion coefficient is
\begin{equation}
D_{e}= m \gamma_{e}\Omega \coth \frac{\Omega }{2T_{e}}
\label{p5d}
\end{equation} where $\Omega=\hbar\omega_{0}$ and the voltage
dependent parameter, $T_{e}$, is given by the relation
\begin{equation}
\coth \frac{\Omega }{2T_{e}}=
\frac{V+\Omega}{2 \Omega}\coth\frac{V+\Omega}{2T}+\frac{V-\Omega}{2 \Omega}\coth\frac{V-\Omega}{2T}
\;.
\label{sw}
\end{equation}
Here $T$ is the temperature of the junction and we assume a {\em dc}
voltage bias, $V= eU$, $U$ being the applied voltage.  We note that
the r.h.s. of Eq.~(\ref{sw}) is proportional to the well-known value
of the power spectrum of current noise of the isolated junction, taken
at the frequency of the oscillator.\cite{Kog96} The fact that the
junction as a part of the environment is in a non-equilibrium state,
is reflected in the voltage dependence of the electronic contribution
to the diffusion coefficient.  In section \ref{stationary} we show
that $T_{e}$ is the effective temperature of the junction seen by the
oscillator.

The phenomenological parameters $D_{0}$ and $\gamma_{0}$ are related
to each other by virtue of the fluctuation-dissipation theorem.
Assuming that the junction and the part of the environment responsible
for $\gamma_{0}$ and $D_{0}$ have the same temperature, $T$, the
diffusion coefficient can be generally presented in the form
\begin{equation}
D = m \Omega \, \left(\gamma_{0}\,\coth \frac{\Omega }{2T} +
\gamma_{e}\,\coth \frac{\Omega }{2T_{e}}
\right)\;.  
\label{vae} 
\end{equation}

The master equation Eq.~(\ref{meqfin}) contains the term proportional
to the coupling constant $A$. A term with this structure has been
obtained in previous discussions of quantum Brownian
motion.\cite{HaaRei85,BrePet02} The derivation of the master equation
for the oscillator (see appendix \ref{XConditional master equation})
shows that the main contribution to the coefficient $A$ in
Eq.~(\ref{meqfin}) comes from virtual tunneling processes with energy
difference of initial and final states of the order of the Fermi
energy, in contrast to the friction and diffusion coefficients which
are controlled solely by the tunneling events in the vicinity of the
Fermi surface.  Besides, compared with the other terms in
Eq.~(\ref{meqfin}), the $A$-term has different symmetry relative to
time reversal, i.e., the transformation $\hat{\rho }\rightarrow
\left(\hat{\rho} \right)^{*}$ and $t \rightarrow -t$.  The damping and
diffusion terms, which are odd relative to time reversal, describe the
irreversible dynamics of the oscillator, whereas the last term in
Eq.~(\ref{meqfin}) just like the Hamiltonian term is time reversible.
These observations give the hint that the $A$-term is responsible for
renormalization-like effects.  This suggests that the $A$-term should
be treated on a different footing than the dissipative terms.
Indirectly, the $A$-term can be absorbed into the Hamiltonian dynamics
at the price of having the time evolution of the oscillator described
by a ``renormalized'' density matrix.  Indeed, if we apply a
(non-unitary) transformation to the density matrix,
$\hat{\tilde{\rho}}=\mathcal{R}\left\{ \hat{\rho}\right\} $, by acting
on the density matrix $\hat{\rho}$ with the super-operator
\begin{equation}
\mathcal{R}\left\{ \hat{\rho}\right\} 
=\hat{\rho}+\mu[\hat{p},[\hat{p},\hat{\rho}],
\label{57d}
\end{equation}
we can by proper choice of the parameter $\mu$,
\begin{equation}
\mu=\frac{A}{2m\Omega^{2}},
\label{ia2}
\end{equation}
cancel the $A$-term in the equation for the transformed matrix
$\hat{\tilde{\rho}}$.  Leaving the $\gamma $ and $D-$terms intact, the
counter term is produced by the super-operator $\mathcal{R}$ acting on
the Hamiltonian part of the master equation Eq.~(\ref{meqfin}).
Applied to the original $A$-term, this procedure generates an
additional contribution proportional to the product $\mu A\propto
A^2$, and it can be neglected to lowest order in the coupling, the
limit which can be consistently studied.

The renormalized density matrix now obeys the master equation
\begin{equation}
\dot{\hat{\rho}}(t)=\mathcal{K}\left\{ \hat{\rho}\right\}
\label{meqfinfin}
\end{equation}
where
\begin{equation}
\mathcal{K}\left\{ \hat{\rho}\right\}
=\frac{1}{i\hbar}[\hat{H}_{R},\hat{\rho}]
+\frac{\gamma}{i\hbar}\left[\hat{x},\left\{
    \hat{p},\hat{\rho}\right\}
\right]-\frac{D}{\hbar^{2}}[\hat{x},[\hat{x},\hat{\rho}]],
\label{}
\end{equation}
up to a term quadratic in the coupling constant. For compact notation,
we have introduced the super-operator $\mathcal{K}$, and dropped the
tilde for marking the renormalized density matrix: thus in the
following the renormalized density matrix will also be denoted by
$\hat{\rho}$.  The master equation being derived for the case of
coupling to a tunnel junction is seen to be of the same {\em form} as
for coupling linearly to a heat bath, i.e., an equilibrium state of a
many-body system; the generic form of a damped quantum oscillator
known from numerous investigations on quantum Brownian
motion.\cite{CalLeg83} However, the diffusion term is qualitatively
different from the usual case where the quantum object is coupled only
to a heat bath. The non-equilibrium state of the junction,
characterized by its voltage, gives rise to features not present when
the coupling is simply to a many-body system in equilibrium.

We note that the super-operator, $\mathcal{R}$, does not change the
trace of the density matrix it operates on, and the renormalized
density matrix is also normalized to unity. However, one has to keep in mind
that the observables are to be calculated with the ''unrenormalized''
density matrix. Up to first order in the coupling constant the
expectation value of an observable $O$ are now given in terms of the
renormalized density matrix according to
\begin{equation}
\langle
O\rangle=\textnormal{Tr}\left(\left(\hat{O}
-\frac{A}{2m\Omega^{2}}
[\hat{p},[\hat{p},\hat{O}]]\right)\hat{\rho}\right).
\label{ja2}
\end{equation}
This relation transfers renormalization from the density matrix to
 observables. In the language of the Feynman diagram technique,
Eq.~(\ref{ja2}) corresponds to a vertex correction.

In this paper, we use the Markovian approximation to describe the time
evolution of the density matrix.  This approximation is valid in the
low frequency limit.  For the {\em dc}-bias case, the characteristic
frequency of time variation of the density matrix, $\omega $, must be
small enough to meet the condition
\begin{equation}
 \omega \ll \omega_{max}
\quad,\quad \omega_{max}\equiv
{\rm max}
\left(\frac{T}{\hbar }, \frac{V}{\hbar } \right).
\label{nbe}
\end{equation}
From the unconditional master equation, the characteristic frequency
is seen to be determined by the friction coefficient, $\omega \sim
\gamma $. This means that the coupling constant $G_{xx}$ in
Eq.~(\ref{gamma}) must be small enough to meet the condition $\gamma
\ll \omega_{max}$.

\section{Current-voltage characteristic}
\label{Current-voltage characteristic}

The average value of the current through the junction is given in
terms of the probability distribution for charge transfers,
i.e.,\cite{RamSheWab04}
\begin{equation}
I(t)=-e\frac{d}{dt}\sum_{n}n\textrm{Tr}\hat{\rho}_{n}(t)
\label{t5d}
\end{equation}
where $\textrm{Tr}$ denotes the trace with respect to the degree of
freedom of the coupled quantum object.  However, to lowest order in
the tunneling, the average current turns out to be expressible through
the charge unconditional density matrix, the reduced density matrix
for the coupled quantum object. Indeed, the master equation for the
charge specific density matrix then enables one to express the time
derivative in Eq.~(\ref{t5d}) in terms of the reduced density matrix
for the quantum object, the charge unconditional density matrix
\begin{equation}
I(t)=e \, \textrm{Tr}\, \mathcal{J}\{\hat{\rho}(t)\}
\label{cur}
\end{equation}
where the drift super-operator, $\mathcal{J}$, is specified in
Eq.~(\ref{2drift2}).

\subsection{Contributions to the current under {\em dc} bias}

 For a {\em dc} bias $V= eU$, $U$ being the applied voltage, the drift
operator $\mathcal{J}$ is given by Eq.~(\ref{2drift2}), and the
current, Eq.~(\ref{cur}), is specified by
\begin{equation}
\frac{1}{e}
I(t)=V\langle G\rangle_{t}+I_{q}(V)
+I_{p}(t)
\label{ifluc}
\end{equation}
and in general time dependent due to the coupling to the oscillator.
The current consists of three physically distinct contributions.  The
first term is the Ohmic-like part of the current proportional to the
conductance
\begin{equation}
\langle G\rangle_{t}= G_{0}+2 G_{x}\langle x\rangle_{t}+ G_{xx}\langle x^{2}\rangle_{t}\quad,
\label{u5d}
\end{equation}
the instantaneous value of the conductance operator,
Eq.~(\ref{w5d}), where $\langle x^{n}\rangle_{t}\equiv\textnormal{Tr}\left(\hat{x}^{n}\hat{\rho}(t)\right)$. We note that besides the pure Ohmic term of the
isolated junction, the additional terms due to the coupling to the
oscillator will in general contribute to the non-linear part of the
current-voltage characteristic since the state of the oscillator will
depend on the voltage. A case in question is discussed in the next
section, where the stationary state of the oscillator is considered.

The second term, $I_{q}$, originates according to the derivation due to
position and momentum operators are not commuting and for this reason we
refer to it as the quantum correction to the current
\begin{equation}
I_{q}(V)=
-\frac{1}{2} 
\tilde{G}_{xx}\,
\Delta_{V}
\; ,
\label{nlin}
\end{equation}
where $\Delta_{V}$ is specified in Eq.~(\ref{dv}) or equivalently
\begin{equation}
\Delta_{V}=
V \; + \; 
(\Omega + V) N_{\Omega + V}
\; - \; 
(\Omega - V) N_{\Omega - V}
\label{8c2}
\end{equation}
and $N_{\Omega \pm V} = 1/(e^{\frac{\Omega \pm V }{T}}-1)$.

The last term in Eq.~(\ref{ifluc}),
\begin{equation}
I_{p}(t)=e\hbar g_{x}\langle\dot{\hat{x}}\rangle_{t}\quad,
\quad\dot{\hat{x}}=\frac{\hat{p}}{m} \;,
\label{9c2}
\end{equation}
is proportional to the average velocity of the oscillator, and present
only for an asymmetric junction, $g_{x}\neq 0$.

The Ohmic part of the current is calculated in section \ref{$I-V$
characteristic} for the stationary case. Next we turn to discuss the
quantum correction and the dissipationless contribution to the current.

\subsection{Quantum correction to the current}

\begin{figure}
\includegraphics[width=0.9\columnwidth]{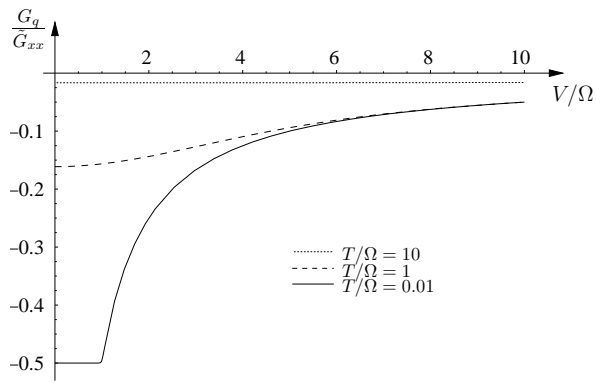}
\caption{
\label{fifluc} 
Voltage dependence of the quantum correction to the junction
conductance, $G_{q}$.  The three curves correspond to the following
relations between the junction temperature $T$ and the oscillator
frequency: $T=0.01\Omega$, $T=\Omega$, and $T=10\Omega$.  At high
temperatures, $T\gg\Omega$, the quantum correction is small and
shows no particular voltage dependence (dotted line). At low
temperatures, $T\ll\Omega$, two distinct regions can be
distinguished. At voltages lower than the oscillator frequency the
quantum correction stays constant, and at voltages higher than the
oscillator frequency it is  inversely proportional to the voltage.  }
\end{figure}

The quantum dynamics of the oscillator leads to a suppresion of the
{\em dc} current as expressed by the quantum correction to the
current, $I_{q}(V)$. Unlike the other terms in the expression for the
current, Eq.~(\ref{ifluc}), the quantum correction, $I_{q}(V)$, does
not depend on the state of the oscillator, but only on its
characteristic energy and the temperature of the junction.  At low
voltages, $V\ll T$, the quantum correction
is linear in the voltage
\begin{equation}
I_{q} = -  V
\tilde{G}_{xx}
\left(
\frac{1}{2} +  N_{\Omega}
 - \frac{\Omega}{T}
N_{\Omega}(N_{\Omega} + 1) \right)
\label{ad2}
\end{equation}
where $N_{\Omega} = 1/(e^{\frac{\Omega }{T}}-1)$.  At large voltages
it reaches a constant value
\begin{equation}
I_{q}\approx
-
\frac{1}{2 }
\tilde{G}_{xx}\,\Omega  
\quad,\quad V\gg T, \Omega  \;,
\label{bd2}
\end{equation} 
in agreement with an earlier result obtained by a technique valid at
zero-temperature.\cite{MozMar02} Our approach generalizes the
expression for the current to arbitrary relations between junction
voltage and temperature, and the frequency of the oscillator.  The
voltage dependence of the quantum correction to the
 conductance, $G_{q}=I_{q}/V$, is
shown in Fig.\ref{fifluc} for different temperatures.

\subsection{Dissipationless current}

The last contribution in Eq.~(\ref{ifluc}) to the current, $I_{p}$, is
qualitatively different from the other terms.  From Eq.~(\ref{9c2}),
one sees that $I_{p}$ is proportional to the velocity of the
oscillator. Therefore, the corresponding transferred charge through
the junction, $\delta Q_{p}=I_{p}\delta t$, is controlled by the
coordinate of the oscillator: $\delta Q_{p}=e \hbar
g_{x}\;\delta\langle x\rangle$.  Being proportional to a velocity, the
current contribution $I_{p}$ is odd with respect to time reversal, and
therefore a dissipationless current.

The presence of the term $I_{p}$ in the current, which does not depend
explicitly on the voltage, means that a current through the junction
can be induced just by the motion of the oscillator alone, i.e., by
the time variation of a system parameter which in our case is the
junction transparency. This effect is closely related to the well
known physics of quantum pumping,\cite{Bro98,AltGla99} but has not, to our
knowledge, been discussed in the present context.  The dissipationless
``pumping''-like current, $I_{p}$, is proportional to the coupling
constant $ g_{x}$.  This contribution to the current is thus only
present if the tunnel junction is asymmetric.  This is in concordance
with the quantum pumping effect.\cite{AleAltKam00}
The global symmetry properties of the system thus crucially determine
the existence and magnitude of the induced current.

A single mode oscillator driven by an external periodic force at
frequency $\omega $ induces an $ac$-current, $I_{\sim}$, with the same
frequency and a phase of the {\em ac} current rigidly following the
phase of the external force.  For a given amplitude of the
oscillations, $x_{max}$, the magnitude of the $ac$-current can be
estimated as
\begin{equation}
I_{\sim}\sim \alpha_{as} \,e \,\omega
\label{cd2}
\end{equation}
where the dimensionless parameter $\alpha_{as}= \hbar g_{x}x_{max}$
characterizes the effective asymmetry of the junction.  In principle,
$\alpha_{as}$ may be comparable to unity so that $I_{\sim} \sim e
\omega $, provided the amplitude of the oscillations $x_{max}$ is
large enough and the conductance of the junction is not too small.

One can show that in the case of an oscillator with two or more modes
interacting with the junction, the corresponding term generates
directed pumping of charge.\cite{denis}

\section{Stationary state properties } \label{stationary}

In this section, we shall study the stationary state of the reduced
density matrix for the oscillator in the Markovian approximation.  
The question arises whether the stationary state of the oscillator is a
thermal equilibrium state even though the environment is in a 
non-equilibrium state as the junction is biased. 
According to Eq.~(\ref{meqfinfin}), the
stationary renormalized density matrix of the oscillator,
$\hat{\rho}_{s}$, is determined by the equation
\begin{equation}
\mathcal{K}\left\{ \hat{\rho}_{s}\right\} =0,
\end{equation}
and the solution is indeed of the form of a thermal density matrix,
$\hat{\rho}_{s} \propto \exp\left(-\hat{H}_{R}/T^{*}\right)$, where
the temperature of the oscillator, $T^{*}$, is specified by the
relation $$\coth\frac{\Omega}{2T^{*}}=\frac{D}{\gamma
m\Omega}.$$ Using Eq.~(\ref{vae}), the temperature of the
oscillator is related to the environment temperature and the voltage
bias according to
\begin{equation}
 \coth\frac{\Omega}{2T^{*}}= 
\frac{\gamma_{0}}{\gamma_{0}+ \gamma_{e}}
\,\coth \frac{\Omega }{2T} +
\frac{\gamma_{e}}{\gamma_{0}+ \gamma_{e}}\,\coth \frac{\Omega }{2T_{e}}.
\label{fd2}
\end{equation}
The average occupation number for the oscillator, $N^{*}$, is given by
the Bose function, $N^{*}= 1/(e^{\Omega /T^{*}}-1)$, and seen to be
populated separately by the interaction with the two environments
\begin{equation}
N^{*}= 
\frac{\gamma_{0}}{\gamma}N_{\Omega}
+
\frac{\gamma_{e}}{\gamma}N_{e}
\label{qbe}
\end{equation}
where $N_{e} = 1/(e^{\frac{\Omega }{T_e}}-1)$.\cite{virial}

We observe, that the oscillator acquires the temperature of the bath,
$T^{*}\approx T$, if the interaction with the junction is weak and
$\gamma_{0}$ is the dominant contribution to the friction,
$\gamma_{0}\gg \gamma_{e}$.  The general case and the opposite limit
where the dynamics of the junction is dominating, $\gamma_{e}\gg
\gamma_{0}$, we proceed to consider.

\subsection{Oscillator heating}\label{}

When the oscillator is well isolated and the interaction with the
junction dominates, $\gamma_{e}\gg \gamma_{0}$, the oscillator attains
according to Eq.~(\ref{fd2}) the effective temperature of the junction
$T_{e}$, as given by Eq.~(\ref{sw}).  As expected, in the absence of a
bias voltage across the junction, $V=0$, the temperature of the
oscillator equals that of the junction irrespective of its
temperature.  When the junction is biased, the oscillator is generally
heated except at zero temperature and low voltages, and we first
discuss the case of a junction at zero temperature.

At zero junction temperature, $T=0$, we must distinguish two voltage
regions.  If the voltage is smaller than the frequency of the
oscillator, $V < \Omega$, the temperature of the oscillator is also
zero, $T^{*}=0$, independent of the voltage as it follows from
Eq.~(\ref{fd2}).  In this regime, the interaction with the tunneling
electrons is unable to excite the oscillator from its ground state.
Heating can only take place beyond the voltage threshold given by the
oscillator frequency.  If instead the voltage is larger than the
frequency of the oscillator, $V > \Omega$, the temperature of the
oscillator, $T^{*}$, is determined by the following relation to the
voltage
\begin{equation}
\tanh \frac{\Omega}{2T^{*}} =\frac{\Omega}{|V|},
\label{gd2}
\end{equation}
At high voltages, $V\gg\Omega$, the temperature of the oscillator
approaches half the bias voltage, $T^{*}=V/2$, in agreement with the
result obtained in a previous study where the temperature of the
junction was assumed to vanish.\cite{MozMar02}

\begin{figure}
\includegraphics[width=0.9\columnwidth]{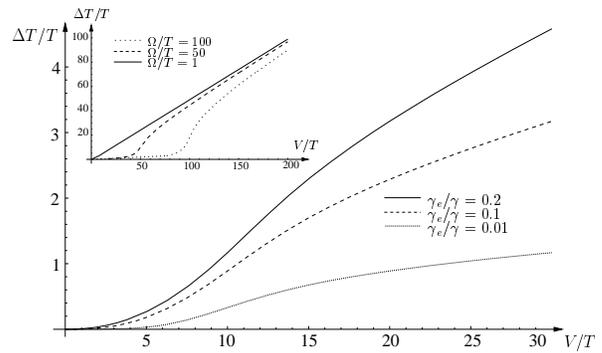}
\caption{
\label{teff}
Heating of the oscillator, its excess temperature, $\Delta T=T^{*}-T$,
as a function of the bias voltage for the temperature $T=0.1 \Omega$,
and different ratios of $\gamma_{e}/\gamma$, covering from moderate to
strong external coupling.  The influence of the external damping
reduces the heating effect. The inset shows the heating with
negligible external damping ($\gamma_{e}/\gamma=1$) as a function of
the bias voltage for different junction temperatures.  At junction
temperatures low compared to the oscillator frequency, $T\ll\Omega$,
two regions of voltage dependence can be distinguished with a rapid
switch as the voltage passes the value of the frequency of the
oscillator. If the bias voltage is smaller than the oscillator
frequency, $V<\Omega$, the oscillator temperature, $T^{*}$, depends
only weakly on the voltage. In the region where the voltage exceeds
the oscillator frequency, $V>\Omega$, the oscillator temperature
approaches $V/2$.  } 
\end{figure}

The heating of the oscillator, its excess temperature, $\Delta T=T^{*}
-T$, as a function of the bias, is shown in Fig.~\ref{teff}, 
both for the case where the coupling to the junction
dominates and the opposite case of dominating external damping.  The
effect of the external damping is shown for moderate to strong
external coupling, $\gamma/\gamma_{e} = 5, 10, 100$.  Increasing the
coupling to the external heat bath leads to suppression of the heating
of the oscillator, the additional environment acting as a heat sink.
In the case where the coupling to the junction dominates, the inset
shows that at low temperatures the oscillator is not excited at
voltages below the frequency of the oscillator.

At high voltages, $V\gg\Omega,T$, in the shot noise regime, the
temperature of the oscillator $T^{*}$ can be found from
Eq.~(\ref{gd2}). Just as in the case of vanishing junction
temperature, the oscillator temperature approaches half the bias
voltage, $T^{*}=V/2$, at large bias, and these results generalize
previous studies which were limited to zero temperature and high
voltages, $V\gg\Omega$.\cite{MozMar02,CleGir04}

In the quest for using tunnel junctions to measure the position of a
coupled object with the ultimate precision set by the uncertainty
principle,\cite{KnoCle03,LahBuuCam04,GaiZolBad05,CleAldDri02} it is
important to take into account that the measuring, involving a finite
voltage, will invariably heat the oscillator.  In this respect the
presence of the additional heat bath, described by the coupling
$\gamma_{0}$, is important. For example, envisioning the oscillator
has been cooled to a temperature $T$ much lower than $\Omega$ and a
voltage is turned on. In order to obtain an appreciable signal the
voltage must be larger than $\Omega$. The oscillator will then in a
time span of the order of $\gamma^{-1}$ be heated and attain a
temperature for which the average number of quanta in the oscillator
is 
\begin{equation}
N^{*} = \frac{\gamma_{e}}{2 \gamma} \frac{|V| -\Omega}{\Omega } .
\label{}
\end{equation}
Estimating the
oscillator temperature, we have $T^{*} \sim
\Omega/\ln(\gamma/\gamma_{e})$.  A strong environmental coupling can
thus be beneficial for retaining the oscillator in the ground state.

\subsection{$I-V$ characteristic}
\label{$I-V$ characteristic}

In the stationary state, the {\em dc} current $I(V)$,
Eq.~(\ref{ifluc}), is conveniently written on the form
\begin{eqnarray}
\frac{1}{e}I(V) = VG_{0} &+&  \frac{1}{2}\tilde{G}_{xx}
\left(
\rule[-1ex]{0ex}{0ex}\right.
2V N^{*}  +
(\Omega -V)N_{\Omega -V} \nonumber \\
&& \quad \quad \quad -
\left.(\Omega +V) N_{\Omega +V}
 \rule[-1ex]{0ex}{0ex} 
\right) \; .
\label{hd2}
\end{eqnarray}
As before, $N^{*}$ is the occupation number 
of the oscillator at the temperature $T^{*}$.
In the
stationary state, only the coupling constant $G_{xx}$ is effective.

The first term in Eq.~(\ref{hd2}) is the current
through the isolated junction, and the remaining term describes 
the influence of the oscillator 
on the current.  
It is interesting to consider the latter in the limit of
vanishing environment temperature, $T=0$.  At zero temperature, we
must distinguish two voltage regimes.  If the voltage is smaller than
the frequency of the oscillator, $V < \Omega$, we observed in the
previous section that the oscillator remains in the ground state,
$N^{*}=0$, and the average current turns out to be equal to that of an
isolated junction, $I = eG_{0}V $. 
One observes that the effect of zero point fluctuations present in the
conductance, Eq.~(\ref{u5d}), is exactly cancelled by the quantum
correction,  Eq.~(\ref{nlin}). 
This lack of influence of zero point fluctuations is expected
since the oscillator in its ground state is inert to the tunneling
electrons for such low bias. 

If the voltage is larger than the oscillator frequency, $V > \Omega$,
the slope of the I-V characteristic, at $T=0$, abruptly
increases, $I = V G_{0}+ (V - \Omega)\tilde{G}_{xx}/2$.

At arbitrary temperatures, the linear conductance of the junction, $G=
I/V$, $V \rightarrow 0$, is given by
\begin{equation}
\frac{1}{e^2}G = G_{0} +\tilde{G}_{xx}
\frac{\Omega }{T}\; N_{\Omega }(N_{\Omega }+1) 
\; .
\label{jd2}
\end{equation}
To derive this formula, we recall that in the limit of vanishing bias,
$V \rightarrow 0$, the oscillator attains the temperature of the junction,
$T^{*}=T$.

At high temperatures, $T\gg V,\,\Omega$, the quantum correction to
the current, and thereby the nonlinear quantum corrections, vanish,
and we obtain the result
\begin{equation}
\frac{1}{e}I(V)=V\left(G_{0}+ G_{xx} \langle x^2\rangle_{*}\right)
\quad,\quad 
\langle x^2\rangle_{*}= \frac{T^{*}}{m \omega_{0}^2}
\end{equation}
where $\langle x^2\rangle_{*}$ is the mean square of the oscillator
coordinate at temperature $T^{*}$.  This result is to be expected from
a classical oscillator in thermal equilibrium influencing the
conductance of a tunnel junction. The $I-V$ characteristic is in this
regime nonlinear due to the voltage dependence of $T^{*}$,
Eq.~(\ref{fd2}).

\section{Current noise} \label{noise}

In this section we use the charge projection technique to
develop the description of the statistical properties of the current
of a tunnel junction coupled to a quantum object. The discussion will
be kept quite general before we in the next section specialize to the
case of a harmonic oscillator, the nano-electromechanical model of
section \ref{REDUCED MASTER EQUATION}.  We shall show how the charge
dynamics, described by the master equation for the charge specific
density matrix, can be used to obtain the statistical properties of
the junction current, such as the noise power spectrum.  The
prerequisite for the success of this endeavour is that for the
considered low transparency tunnel junction, the charge representation
in fact provides the probability distribution for the charges {\em
transferred} through the junction.\cite{SheRam03,RamSheWab04}

\subsection{ Current noise in the charge representation}

The probability, $p_{n}(t)$, for $n$ charge-transfers in time span $t$
is according to Eq.~(\ref{jsd}) given by
$p_{n}(t)=\textnormal{Tr}\,\hat{\rho}_{n}(t)$, where
$\hat{\rho}_{n}(t)$ is the charge specific density matrix,
Eq.~(\ref{Y10sd}).  The charge-transfer probability distribution
specifies the stochastic process of charge transfers, $n(t)$.  The
variance of the charge fluctuations
\begin{equation}
\left\langle \left\langle n^{2}(t)\right\rangle \right\rangle
=\left\langle n^{2}(t)\right\rangle -\left\langle n(t)\right\rangle
^{2},
\end{equation}
is defined in terms of the moments of the probability distribution of
charge transfers
\begin{equation}
\left\langle n^{r}(t)\right\rangle
=\sum_{n}n^{r}\,p_{n}(t)
\;,\; 
p_{n}(t)=
\textnormal{Tr}\left(\hat{\rho}_{n}(t)\right)
\;
.
\end{equation}

To express the statistical properties of the current in terms of the
probabilities for charge transfers, we inherit the stochastic current
process, $i(t)$, through its relation to the charge transfer process
\begin{equation}
n(t)=\int_{0}^{t}dt'i(t').
\label{h8d}
\end{equation}
The average current, given by $\left\langle i(t)\right\rangle =
d\left\langle n(t) \right\rangle/dt $, is in accordance with
Eq.~(\ref{t5d}).  The variance of the charge fluctuations are
expressed via the current fluctuations according to
\begin{equation}
\left\langle \left\langle n^{2}(t)\right\rangle \right\rangle
=\int_{0}^{t}dt_{1}\int_{0}^{t} dt_{2}\,\langle\delta
i(t_{1})\delta i(t_{2})\rangle\;.
\label{m9d}
\end{equation}
where $\delta i(t)=i(t)-\left\langle i\right\rangle$, and $\langle
i\rangle$ is the average $dc$ current since we in the following shall
consider the stationary state.  Stationary current noise is
characterized by the current-current correlator
\begin{equation}
S(\tau)=\left\langle \delta i(t+\tau)\delta i(t)\right\rangle
\;.
\label{m8d}
\end{equation}
Inserting this expression into Eq.~(\ref{m9d}) one obtains
\begin{equation}
\left\langle \left\langle n^{2}(t)\right\rangle \right\rangle
=2\int_{0}^{t}d\tau\,(t-\tau)S(\tau)\,.
\label{n9d}
\end{equation}
This expression allows one to relate the charge and current
fluctuations.

Taking time derivatives of Eq.~(\ref{n9d}) gives
\begin{equation}
\frac{d\left\langle \left\langle n^{2}(t)\right\rangle \right\rangle
}{dt}=2\int_{0}^{t}d\tau\, S(\tau),
\label{j8d}
\end{equation}
and
\begin{equation}
S(t)=
\frac{1}{2}
\frac{d^{2}\left\langle \left\langle n^{2}(t)\right\rangle
  \right\rangle }{dt^{2}},
\label{i8d}
\end{equation}
i.e., the current-current correlator equals the second derivative of
the variance of charge transfers.  This relation allows one to
calculate the current-current correlator, $S(t)$, by evaluating the
charge fluctuations using the master equation for the charge specific
density matrix.

Eventually, interest is in the current noise power spectrum,
$S_{\omega}$, given by
\begin{equation}
S_{\omega}=4\int_{0}^{\infty}dt \,\cos(\omega t)S(t)
\label{kd2}
\end{equation}
where $\omega$ is the frequency at which the noise is
measured.\cite{Ric54}

We observe that the zero frequency noise power, according to
Eqs.~(\ref{j8d}) and (\ref{kd2}), can be calculated from the general
relation
\begin{equation}
S_{\omega=0}=
2
\left.\frac{d\left\langle \left\langle n^{2}(t)
\right\rangle \right\rangle}{dt}
\right|_{t \rightarrow \infty }
\label{nae}
\end{equation}
i.e., as the rate of change of the charge variance at large times.

In the present approach it is convenient to calculate directly the
current-current correlator, as done in the next section and appendix
\ref{tech}. Only at the end we then transform to obtain the noise
power spectrum. However, we note that the approach is equivalent to
employing the widely used MacDonald formula.\cite{MacD}

\subsection{Current-current correlator} \label{curcur}

In this section we show how the master equation for the charge
specific density matrix, can be used to obtain the noise power
spectrum of the current.  A convenient feature of the method is that
it allows one directly to obtain the time dependence of the current noise.

The probability distribution of charge transfers, $p_n(t)$, is
obtained from the master equation for the charge specific density
matrix given the initial condition corresponding to a state of
definite initial charge
\begin{equation}
\hat{\rho}_{n}(t=0)=\delta_{n,0}\,\hat{\rho}_{s}
\label{s8d}
\end{equation}
at the time when the charge counting starts, the initial time $t=0$.
We are interested in the noise properties of the stationary state and,
therefore, the stationary density matrix of the oscillator, the
thermal state $\hat{\rho}_{s}$, enters the initial condition.

In the following we shall treat the charge specific dynamics 
in the Markovian approximation.  
The charge specific density matrix $\hat{\rho}_{n}(t)$, is
obtained as the solution of the master equation Eq.~(\ref{conmeq}).
For notational convenience we write the charge specific master
equation in the form
\begin{equation}
\frac{d\hat{\rho}_{n}}{dt}=\mathcal{K}\left\{ \hat{\rho}_{n}\right\}
+\mathcal{D}\left\{ \hat{\rho}_{n}''\right\} +\mathcal{J}\left\{
  \hat{\rho}_{n}'\right\} ,
\label{k9d}
\end{equation}
where $\mathcal{K}$ is the super-operator introduced in
Eq.~(\ref{meqfinfin}). Although the $\hat{\rho}_{n}$'s are time
dependent, the unconditional density matrix,
$\hat{\rho}=\sum_{n}\hat{\rho}_{n}(t)$, remains equal to the thermal
state, $\hat{\rho}_{s}$, by virtue of its stationarity property,
$\mathcal{K}\left\{ \hat{\rho}_{s}\right\}=0$.  Our goal is now to
evaluate the variance of the charge transfers and thereby the
current-current correlator with the help of Eq.~(\ref{i8d}).

The rate of change of the first charge moment, {\it i.e.}, the {\em dc}
current according to Eq.~(\ref{t5d}), becomes in the stationary state
\begin{equation}
\frac{1}{e}I
=-\textnormal{Tr}\left(\mathcal{J}\left\{ \hat{\rho}_{s}\right\}
\right),
\label{current}
\end{equation}
following from Eq.~(\ref{cur}) and the stationarity property of the
density matrix for the coupled quantum object, $\hat{\rho}(t)=
\hat{\rho}_{s}$.
The {\em dc} current was calculated in section \ref{$I-V$ characteristic}.

It readily follows Eq.~(\ref{k9d}), that the time derivative of the 
variance of charge
transfers, $\left\langle \left\langle n^{2}(t)\right\rangle
\right\rangle$, can be presented in the form
\begin{equation}
\frac{d}{dt}\left\langle \left\langle n^{2}(t)\right\rangle
\right\rangle =2\textnormal{Tr}\left(\mathcal{D}\left\{
    \hat{\rho}_{s}\right\}
\right)-2\textnormal{Tr}\left(\mathcal{J}\left\{
    \delta\hat{N}(t)\right\} \right),
\label{deriv_variance}
\end{equation}
 where $\delta\hat{N}(t)$ denotes the traceless matrix
\begin{equation}
\delta\hat{N}(t)=\sum_{n}\left(n-\left\langle n(t)\right\rangle
\right)\hat{\rho}_{n}(t).
\label{b}
\end{equation}
We observe that only the truncated density matrix, $\delta
\hat{N}(t)$, is needed to calculate the noise.

Comparing Eq.~(\ref{deriv_variance}) and Eq.~(\ref{j8d}), one
concludes that the current-current correlator has a $\delta -$function
like singularity at the initial time, $t=0$, where the charge counting
starts. Indeed, the r.h.s. of Eq.~(\ref{deriv_variance}) has a finite
limit as $t \rightarrow 0$, given by the first term, since the second
term initially vanishes, $\delta \hat{N}(t=0)=0$.  For this result to
be compatible with Eq.~(\ref{j8d}) and Eq.~(\ref{i8d}), the
current-current correlator, $S(t)$, must have the following structure
\begin{equation}
S(t) = S_{1}(t) + S_{2}(t) \,,
\label{p9d}
\end{equation}
the sum of a singular contribution,
\begin{equation}
S_{1}(t) = 
2\textnormal{Tr}\left(\mathcal{D}\left\{
    \hat{\rho}_{s}\right\}\right)
\,
\delta(t) 
\label{q9d}
\end{equation}
where $\delta(t)$ denotes a function peaked at $t=0$ and normalized
according to the condition $\int_{0}^{\infty}dt \;\delta(t)=
\frac{1}{2}$, and a regular part given by
\begin{equation}
S_{2}(t)
=-\textnormal{Tr}\left(\mathcal{J}\left\{
    \delta\hat{I}(t)\right\} \right),
\label{w8d}
\end{equation}
where $\delta\hat{I}$ denotes the matrix, $\delta\hat{I}= \frac{d
}{dt}\,\delta\hat{N}$. The finite time correlation of the current
described by the regular part $S_{2}(t)$ is solely due to the
interaction with the quantum object, as follows from
$\delta\hat{I}(t)$ being traceless.  
We note here that the $\delta $-function singularity,
which would provide noise at arbitrary high frequencies, is an
artefact of the Markovian approximation.

The task of calculating the time dependent current noise is thus
reduced to obtaining the time derivative of the charge-averaged
density matrix, $\delta \hat{N}(t)$, given in Eq.~(\ref{b}).  From the
master equation for the charge specific density matrix one obtains the
following equation for $\delta\hat{I}(t)$,
\begin{equation}
\frac{d}{dt}\delta\hat{I} = \mathcal{K}\{\delta\hat{I}\},
\label{diff_b_1}
\end{equation}
 and the initial condition 
\begin{equation}
\left.\delta\hat{I}\right|_{t=0}=-\delta\mathcal{J}\left\{
\hat{\rho}_{s}\right\} .
\end{equation}
Here the super-operator $\delta\mathcal{J}$ acts on its argument
matrix according to 
\begin{equation}
\delta\mathcal{J}\left\{ X\right\} =\mathcal{J}\left\{ X\right\}
- X\, \left(\textnormal{Tr}\,\mathcal{J}\left\{X\right\} \right).
\label{z8d}
\end{equation}
We note, that acting on a matrix $X$ with unit trace,
$\textnormal{Tr}\,X =1$, the super-operator $\delta {\cal J}$ returns
a traceless matrix.  The dynamics of the charge averaged quantity
$\delta\hat{I}$ is thus identical to that of the charge unconditional
density matrix of the oscillator.

The formal solution to Eq.~(\ref{diff_b_1}) can be written in terms of
the time evolution super-operator for the charge unconditional density
matrix of the oscillator
\begin{equation}
\mathcal{U}_{t}\,
=
e^{\mathcal{K}t}
\label{y8d}
\end{equation}
 as 
\begin{equation}
\delta\hat{I}(t)=
- \,
\mathcal{U}_{t} 
\left\{ 
\delta\mathcal{J}\left\{ \hat{\rho}_{s}\right\} \right\}
\end{equation}
and the regular part of the current-current correlator can be written
on the form
\begin{equation}
S_{2}(t)=\textnormal{Tr}\left(\delta\mathcal{J}\left\{
    \mathcal{U}_{t}\left\{ 
\delta\mathcal{J}\left\{ \hat{\rho}_{s}\right\} \right\} \right\} \right).
\label{S2}
\end{equation}
Here ${\cal J}$ in Eq.~(\ref{w8d}) has been replaced for $\delta {\cal
J}$ in Eq.~(\ref{w8d}); the replacement is valid under the trace
operation since ${\rm Tr}\; \delta \hat{I}(t) =0$.  Combined with the
singular part in Eq.~(\ref{p9d}), this gives the general expression in
the Markovian approximation for the current-current correlator of a
tunnel junction interacting with a quantum system in its stationary
state.  The current noise correlator has thus conveniently been
written with the help of the Markovian super-operators 
${\cal K}$, ${\cal J}$ and ${\cal D}$.

In the next section we shall turn to calculating the noise properties
for the case of the nanoelectromechanical device described in section
\ref{REDUCED MASTER EQUATION}.

\section{ Noise power spectrum }\label{corrr}

We now turn to calculate the current-current correlator of the tunnel
junction coupled to the harmonic oscillator as described by the model 
of section \ref{REDUCED MASTER EQUATION}.  Taking advantage of the general
analysis in the Markovian approximation presented above in Section
\ref{curcur}, the current-current correlator can be written in the form:
\begin{equation}
S(t) = S_{1}(t) + S_{x}(t) + S_{x^2}(t) \; .
\label{ac2}
\end{equation}
Here $S_{1}$ is the singular part defined in Eq.~(\ref{q9d}) and
specified in Eq.~(\ref{pc2}).  
The regular contribution is given by second and third terms on the
right in Eq.~(\ref{ac2}), as obtained by
inserting the expression for the drift super-operator ${\cal J}$, 
Eq.~(\ref{2drift2}), into the expression for the regular 
contribution, Eq.~(\ref{S2}), giving
\begin{equation}
S_{x}(t)= 
2V G_{x}\, x_{J}(t)
+
g_{x}\frac{\hbar }{m}\,  p_{J}(t)
\label{qc2}
\end{equation}
and 
\begin{equation}
S_{x^2}(t)=
VG_{xx}\,  x_{J}^2(t)
\label{rc2}
\end{equation}
where the time dependent quantities are given by
\begin{equation}
X_{J}(t) = 
{\rm Tr}\left(\hat{X} \,\mathcal{U}_{t}\left\{
\delta\mathcal{J}\left\{ \hat{\rho}_{s}\right\} \right\} 
\right),  
\label{sc2}
\end{equation}
with $ \hat{X} = \hat{x}, \hat{p}, \hat{x}^2$, respectively. 
It is readily checked that the latter
quantities evolve in time in accordance with their corresponding
classical equations of motion for a damped oscillator with initial
conditions given by
\begin{equation}
X_{J}(0) = 
{\rm Tr}\left(\hat{X} \,
\delta\mathcal{J}\left\{ \hat{\rho}_{s}\right\} 
\right)
\quad,\quad \hat{X} = \hat{x},\hat{p},\hat{x}^2,\hat{p}^2 
\,.
\label{wc2}
\end{equation}
The calculation of these quantities are presented in appendix
\ref{tech}, giving according to Eq.~(\ref{qc2}) and Eq.~(\ref{rc2}) an
explicit expression for the current-current correlator $S(t)$ and
thereby, according to Eq.~(\ref{kd2}), for the noise power $S_{\omega
}$.  Fourier transforming the current-current correlator gives,
according to Eqs.~(\ref{bla23}, \ref{7b23}, \ref{pc2}),
peaks in the noise power as well as a constant up-shift in the noise
floor, the noise pedestal. The noise power spectrum is displayed in
the inset in Fig.~\ref{asym}. As expected, there is a pronounced
peak at the frequency of the oscillator, and in addition two side
peaks each shifted by the frequency of the oscillator, one at zero
frequency and one at twice the oscillator frequency.

Below, we analyze the noise in two frequency regions: (i) low frequency
noise at frequencies $\omega \lesssim \gamma $; and (ii) noise in the
vicinity of the oscillator resonance frequency $\omega \approx
\omega_{0}$ and $\omega \approx 2 \omega_{0}$.  We examine the voltage
and temperature features of the noise power.

\subsection{Low frequency noise}\label{lowfreq}

Let us consider low frequency noise, at frequencies of the order of
the damping rate and lower, $\omega\lesssim\gamma$.  Then the noise
power spectrum is given by the Fourier transform of the correlation
functions $S_{1}(t)$ and $S_{x^{2}}(t)$ in Eqs.~(\ref{7b23}), and
(\ref{pc2}), respectively, giving
\begin{equation}
S_{\omega}=S^{(0)}+S^{(1)}+S_{\omega}^{(2)}\label{7e2}\end{equation}
 where 
\begin{equation}
S^{(0)}= 2G_{0} V\coth\frac{V}{2T}
\label{ube}
\end{equation}
is the low frequency, $\omega\ll V$, white Nyquist or Schottky noise of
the isolated junction, and $S^{(1)}$ is the correction to the white
noise due to the interaction with the oscillator
\begin{equation}
S^{(1)}=
2\tilde{G}_{xx}\Omega\left(N^{*}(N_{e}+1)+N_{e}(N^{*}+1)\right).
\label{8e2}\end{equation}
Together these two contributions form the noise pedestal, 
$S_{\infty}=S^{(0)}+S^{(1)}$.
The frequency dependent part, $S_{\omega}^{(2)}$, becomes at low
frequencies
\begin{equation}
S_{\omega}^{(2)}=
\tilde{G}_{xx}\frac{4\gamma\gamma_{e}}{\omega^{2}
+4\gamma^{2}}\frac{V}{\Omega}\left(2VN^{*2}+\left(V-
\Delta_{V}\right)\left(2N^{*}+1\right)\right).
\label{9e2}
\end{equation}
The low frequency noise is displayed explicitly proportional to the
coupling to the electronic tunnel junction environment as we have
taken advantage of the relation $\gamma_{e}= \tilde{G}_{xx}\Omega $.
The width of the low frequency peak is twice the damping rate
$2\gamma$.
At zero bias, $V=0$, where $S_{\omega}^{(2)}=0$ and the oscillator and
effective junction temperatures equal the environment temperature, leaving
$N^{*}=N_{e}=N_{\Omega}$,
one recovers the fluctuation-dissipation relation for the noise power, 
$S_{\omega=0}=4TG$,
where $G$ is the linear conductance of the junction in the presence of
the interaction with the oscillator, i.e., given by Eq.~(\ref{jd2}).
 
In the following we discuss the features of the low frequency excess
noise, the noise due to the coupling to the oscillator, and in
particular the noise peak height at zero frequency, in the limits of
temperatures high and low compared to the oscillator frequency.
 
\subsubsection{Low temperature noise }

First we consider the low frequency noise at low temperatures,
$T\ll\Omega$. As expected, no excess
noise is according to Eq.~(\ref{7e2}) generated by the oscillator at
zero temperature and voltages below the oscillator frequency,
$|V|<\Omega$, where the oscillator cannot be excited from its ground
state.  Indeed, in the region of low temperatures, $T \ll \Omega$, and
low voltages, $V < \Omega$, the oscillator is non-responsive and the
excess noise, $S^{(1)}$ and $S_{\omega}^{(2)}$, vanishes exponentially
in $\Omega/T$ below the activation energy $\Omega$.

Close to the noise onset threshold, $|V|\sim \Omega$, in the narrow
region, $||V|-\Omega | \ll T$, the peak height relative to the
pedestal rises linearly with temperature
\begin{equation}
S_{\omega=0}^{(2)}=\frac{\gamma_{e}}{\gamma}T\tilde{G}_{xx}.
\label{if2}
\end{equation}
 
At voltages much higher than threshold, $|V|\gg\Omega$, the peak
height relative to the pedestal becomes
\begin{equation}
S_{\omega=0}^{(2)}=
2\frac{\tilde{G}_{xx}^{2}}{\gamma}V^{2}
\left(
\frac{1}{2} + \frac{\gamma_e}{\gamma}\frac{|V|}{2\Omega}
+
\left(\frac{\gamma_e}{\gamma}\frac{V}{2\Omega} \right)^2
\right).
\label{jf2}
\end{equation}
The zero frequency noise is proportional to $V^{4}$ if the effective
coupling to the electronic environment is appreciable, i.e., the ratio
$\gamma_{e}/\gamma$ is not too small.  In the high voltage limit, the
oscillator is in the classical regime but with the oscillator
temperature given by $T^{*}=|V|/2$ as discussed in section
\ref{stationary}.

\subsubsection{High temperature noise}

At temperatures higher than the oscillator frequency, $T\gg\Omega$, we
can distinguish two voltage regimes.  At low voltages, $V\ll T$, the
peak height scales quadratically in both the temperature and voltage
\begin{equation}
S_{\omega=0}^{(2)}= \frac{2}{\gamma}
\tilde{G}_{xx}^{2}\,
V^{2}
\left(\frac{T}{\Omega}\right)^{2},
\label{peakZerHiT}
\end{equation}
and we recall that the oscillator temperature equals the junction
temperature, $T^{*}=T$.
 
At high voltages, $V\gg T$, the peak height becomes
\begin{equation}
S_{\omega=0}^{(2)}=
\frac{2}{\gamma}V^{2}\tilde{G}_{xx}^{2}\,\left(
\frac{\gamma_{0}}{\gamma}\frac{T}{\Omega}+\frac{\gamma_{e}}{\gamma}\frac{\left|
V\right|}{2\Omega}\right)^{2},
\label{kf2}
\end{equation}

At high temperatures, the oscillator is in the classical
regime and the average occupation number depends linearly on the
oscillator temperature. The peak height, proportional to the
fluctuations in the oscillator position squared, is proportional to
the square of the average occupation number, and is therefore
proportional to the square of the oscillator temperature.

\subsection{High frequency noise}

Next, we investigate the properties of the peaks in the noise power
spectrum occurring at finite frequencies, at the oscillator frequency,
$\omega\approx\omega_{0}$, and its  harmonic,
$\omega\approx2\omega_{0}$.  The Markovian approximation allows us to
consider the high frequency noise only under the condition
Eq.~(\ref{nbe}), that the frequency is much smaller than the maximum
value of the voltage or the temperature, and for frequencies in
question, this means that ${\rm max} (T,V)\gg\Omega$ for consistency.
The inset in Figure \ref{asym} shows the frequency dependence of the
noise power spectrum, Eq.~(\ref{kd2}), in the case of high
temperatures, $T\gg\Omega$. The noise power displays three peaks.  The
noise power spectrum, consists at $\omega\approx\omega_{0}$ of a
Lorentzian part, as given by the Fourier transform of Eq.~(\ref{bla23}), 
with a width given by the damping rate,
$\gamma$, and an asymmetric part specified by Eq.~(\ref{bla23}) and
present only for an asymmetric junction.  At $\omega\approx 2 \omega_{0}$, 
the noise power spectrum is according to Eq.~(\ref{7b23}) a
Lorentzian with a width given by twice the damping rate, $2\gamma$.
We now turn to discuss these peak heights at high and low
temperatures.

\subsubsection{High frequency noise at high temperatures}
 
At high temperatures, $T\gg V\gg\Omega$, the height of the peak at the
oscillator frequency relative to the pedestal depends linearly on
temperature
\begin{equation}
S^{(2)}_{\omega=\omega_{0}}=
2\frac{T}{\gamma}\left[\tilde{G}_{x}^{2}\frac{4V^{2}}{\Omega}-
\tilde{g}_{x}^{2}\Omega\right],
\label{peakOmHiT}
\end{equation}
and is determined by the conductances $\tilde{G}_{x}$ and
$\tilde{g}_{x}$.

At double the oscillator frequency the peak height depends
quadratically on temperature
\begin{equation}
S^{(2)}_{\omega=2\omega_{0}}=\frac{1}{\gamma}\tilde{G}_{xx}^{2}\frac{V^{2}T^{2}}{\Omega^{2}}\label{peak2OmHiT}
\end{equation}
and just as the peak at zero frequency determined by the conductance
$\tilde{G}_{xx}$.  We note that its height is half that of the peak at
zero frequency, Eq.~(\ref{peakZerHiT}).

The expressions for the excess noise, Eq.~(\ref{peakOmHiT}) and
Eq.~(\ref{peak2OmHiT}), are in fact also valid at low voltage.
Contrary to the peaks at $\omega=0$ and $\omega=2\omega_{0}$, which
vanish in the absence of voltage, the excess noise power at $\omega
\approx \omega_{0}$ is therefore finite for an asymmetric junction
even at zero voltage, and according to Eq.~(\ref{peakOmHiT}) in fact
negative.  An asymmetric junction with a $Q$-factor much larger than
$T/\Omega$ can thus at zero voltage lead to a suppression of the noise
power below that of an isolated junction.

\subsubsection{High frequency noise at low temperatures}
 
At high voltages and low temperatures, $V\gg\Omega\gg T$, the peak
height at the oscillator frequency becomes
\begin{equation}
S^{(2)}_{\omega=\omega_{0}}=
\frac{2 V}{\gamma}
\left[2\tilde{G}_{x}^{2}V
\left(1+\frac{\gamma_{e}}{\gamma}\frac{|V|}{\Omega}\right)
-
\tilde{g}_{x}^{2}\Omega\left(1-\frac{\gamma_{e}}{2\gamma}\right)\right],
\end{equation}
and the peak height at twice the oscillator frequency
 \begin{equation}
S^{(2)}_{\omega=2\omega_{0}}
=\frac{1}{\gamma}\tilde{G}_{xx}^{2}V^{2}\,\frac{\gamma_{e}}{\gamma}
\frac{|V|}{2\Omega}\left(\frac{\gamma_{e}}{\gamma}\frac{|V|}{2\Omega}+1\right).
\end{equation}
The noise can be large due to the high oscillator temperature.

\subsubsection{Noise asymmetry}
 
A striking feature of the finite frequency noise is the contribution
proportional to $g_{x}G_{x}$.  It is \emph{odd} relative to the sign
of the voltage and does \emph{not} depend on the state of the
oscillator (see Eq.~(\ref{7b23})).  This term, which is only present
for an asymmetric junction, $g_{x}\neq0$, does not contribute to the
peak height at the oscillator frequency, but provides the asymmetry of
the peak in the frequency region around the oscillator frequency,
$\omega=\omega_{0}$.  Separating the even and odd voltage
contributions in the noise power,
$S_{\omega}^{\pm}=\frac{1}{2}\left(S_{\omega}(V)\pm
S_{\omega}(-V)\right)$, the odd contribution becomes
\begin{eqnarray}
S_{\omega}^{-}&=&\tilde{g}_{x}\tilde{G}_{x}\left(F_{\omega_{0}}(\omega)-F_{-
\omega_{0}}(\omega)\right) \\ &\times& \left(V^{2}\coth\frac{V}{2T}+\Omega
V(2N_{e}+1)-\frac{\Omega}{2}\Delta_{V}\right),\nonumber
\label{Sasym}
\end{eqnarray}
with the frequency dependence given by the function 
\[
F_{\omega_{0}}(\omega)=\frac{2(\omega_{0}-\omega)}
{\gamma^{2}+(\omega-\omega_{0})^{2}}.
\]

\begin{figure}
\includegraphics[width=0.9\columnwidth]{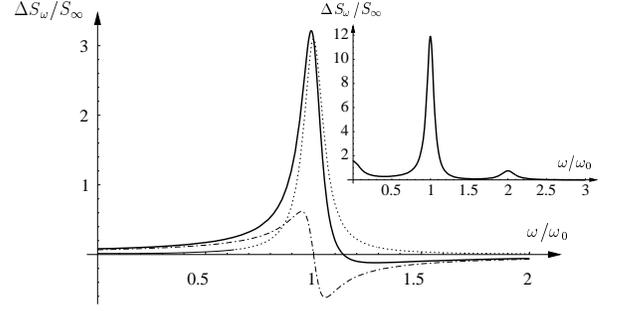}
\caption{\label{asym}The excess noise power spectrum, 
$\Delta S_{\omega}=S_{\omega}-S_{\infty}$, due to an oscillator
coupled to an asymmetric junction is shown by the full line for the
parameter values $V/\Omega=10$, $T/\Omega=0.01$,
$\gamma_{0}/\omega_{0}=0.05$ and conductances $G_{0}=1$,
$\tilde{G}_{x}=0.07$, $\tilde{g}_{x}=0.05$, and $\tilde{G}_{xx}=0.005$
giving for the electronic coupling constant
$\gamma_{e}/\omega_{0}=0.005$.  
The dotted line shows the contribution symmetric
in the voltage, $S_{\omega}^{+}$ ,and the dash-dotted line the
contribution asymmetric in the voltage, $S_{\omega}^{-}$.  The inset
shows the noise power spectrum for the high temperature $T=100\Omega$,
but otherwise the same set of parameters, a regime where the peaks at
zero and twice the oscillator frequency are also visible.}
\end{figure}

The noise power spectrum is displayed by the full line in Figure
\ref{asym} for the temperature $T = 0.01 \Omega$, where only the peak
at the oscillator frequency is appreciable.  The even part in voltage
of the noise power, $S_{\omega}^{+}$, displayed by the dotted line, is
a symmetric function of the frequency relative to the frequency of the
oscillator, and the odd part in the voltage, $S_{\omega}^{-}$,
displayed by the dash-dotted line, is an anti-symmetric function of
the frequency relative to the oscillator frequency.  If the voltage is
reversed, the frequency dependence of the asymmetric part is mirrored
around the frequency $\omega=\omega_{0}$.
 
In contrast to an isolated junction,
the noise power of the coupled junction-oscillator system
shows asymmetry in the voltage. 
This behavior is a novel feature that arises when an asymmetric junction
is coupled to an additional degree of freedom.

\section{Conclusions}

We have applied the charge projection technique to obtain the charge
specific dynamics of a continuous quantum degree of freedom coupled to
a tunnel junction.  The master equation for the charge specific
density matrix has been derived, describing the charge conditioned dynamics
of the coupled object as well as the charge transfer statistics of the
junction.  The method allows evaluating at any moment in time the
joint probability distribution describing the quantum state of the
object and the number of charges transferred through the junction.

The approach, generally valid for any quantum object coupled to the
junction, has been applied to the generic case of a nanoelectromechanical
system, a harmonic oscillator coupled to the charge dynamics of a
tunnel junction.  In this regard it is important that the method
allows inclusion of a thermal environment in addition to the
electronic environment of the tunnel junction since nanoresonators are
invariably coupled to a substrate.  The oscillator dynamics, described
by the reduced density matrix for the harmonic oscillator, the charge
specific density matrix traced with respect to the charge index, has
upon a renormalization been shown to satisfy a master equation of the
generic form valid for coupling to a heat bath.  Even though the
electronic environment is in a non-equilibrium state, the master
equation is of the Caldeira-Leggett type, consisting of a damping and
a fluctuation term. Though the coefficients of the terms are not
related by the equilibrium fluctuation-dissipation relation, the
fluctuation term originating from the coupling to the junction is of
the steady-state fluctuation-dissipation type, containing the current
noise power spectral function of the isolated junction taken at the
frequency of the oscillator.  The diffusion parameter is thus
determined by all energy scales of the problem including temperature,
voltage and oscillator frequency.  The presence of an environment in a
non-equilibrium state thus leads to features which are absent when the
oscillator is only coupled to a heat bath.

The Markovian master equation for the charge specific density matrix
has been used to calculate the current.  In general, the average
junction current consists of an Ohmic term, however, with a
conductance modified due to the coupling to the oscillator dynamics, a
quantum correction, and a
dissipationless {\em ac} current only present for an asymmetric
junction and proportional to the instantaneous velocity of the oscillator.
The latter term does not depend on voltage explicitly and is an
example of an effect similar to quantum pumping.

The stationary state of the oscillator has been shown to be a thermal
state even though the environment is in a non-equilibrium state. 
Therefore, the only effect of the bias is heating of the junction.
Thus the stationary oscillator state is a thermal equilibrium state, 
though in equilibrium at a higher temperature than that of the
environment if the junction is in a non-equilibrium state of finite
voltage. This is a backaction effect of the measuring device, 
the tunnel junction, on the oscillator. 
At zero temperature and voltages below the oscillator
frequency, the oscillator remains, to lowest order in the tunneling, 
in its ground state, and the {\em dc} current equals that of an 
isolated junction.  The coupling of the
oscillator to the additional heat bath, described by the coupling
constant $\gamma_0$, is shown to be beneficial for avoiding heating
of the oscillator due to a finite voltage. This is of importance for
application of quantum point contacts and tunnel junctions to position
measurements aiming at a precision reaching the quantum limit.

The charge projection method has been used to infer the statistical
properties of the junction current from the charge probability
distribution. For example, the noise power spectrum is
specified in terms of the variance of the charge distribution.  
The master equation for the
charge specific density matrix can therefore be used to obtain the
current-current correlator directly, and this has been done explicitly in
the Markovian approximation. The excess noise power spectrum due to
the coupling to the oscillator consists of a main peak located at the
oscillator frequency and two smaller peaks located at zero frequency
and twice the oscillator frequency, respectively.  The peaks at zero
frequency and at twice the oscillator frequency have heights
proportional to the coupling constant $G_{xx}$ squared, whereas the
height of the peak at the oscillator frequency is proportional to the
coupling constants $G_{x}$ and $g_{x}$ squared.  The voltage and
temperature dependencies of the peaks has been examined in detail.

For an asymmetric junction, the noise power spectrum contains a term
with the striking feature of being an odd function of the voltage and
independent of the state of the oscillator.  Contrary to the case of a
symmetric junction, coupling of an oscillator to an asymmetric
junction with temperature higher than the oscillator frequency results
even at zero voltage in a suppression of the noise power at the
oscillator frequency, the excess noise power being negative.  For an
asymmetric junction, the noise power at $\omega \approx \omega_{0}$
can thus be suppressed below the Nyquist level of the isolated
junction.

The Markovian approximation employed to calculate 
the noise power can not be validated at arbitrary frequencies 
compared to temperature or voltage. Not surprisingly, naive attempts to
extend expressions beyond the Markovian applicability range,
Eq.~(\ref{nbe}), 
leads to unphysical results for the noise. For example,
at zero temperature and voltages below the oscillator frequency, 
a spurious noise power arises even for the
oscillator in the ground state.

\acknowledgments
This work was supported by The Swedish Research Council.

\appendix

\section{Charge specific master equation}

\label{XConditional master equation}

In a previous paper we introduced the charge representation for a
general many-body system.\cite{RamSheWab04} The approach is based on
the use of charge projectors previously introduced in the context of
counting statistics.\cite{SheRam03}
In the charge representation, the dynamics of a quantum
object coupled to a many-body system is described by the charge
specific density matrix
\begin{equation}
\hat{\rho}_{n}(t)=\textrm{Tr}_{el}\;(\mathcal{P}_{n}\;\rho(t))
\label{X10sd}
\end{equation}
where $\rho(t)$ is the full density matrix for a many-body system and
a quantum object coupled to it, and $\textrm{Tr}_{el}$ denotes the
trace with respect to the degrees of freedom of the many-body system,
in the following assumed the conduction electrons of a tunnel
junction. The charge projection operators $\mathcal{P}_{n}$, which
project the state of the system onto its component for which
exactly $n$ electrons are in a specified region of space, have been
discussed in detail earlier.\cite{SheRam03,RamSheWab04} 
There we discussed the circumstances
under which the charge index, $n$, can be interpreted as the number of
charges \emph{transferred} through the junction, and the charge
projector method thus provides a basis for charge counting statistics
in the cases where the distribution function for transferred charge is
a relevant concept.  In this case, the charge specific density matrix
allows the evaluation, at any moment in time, of the joint probability
of the quantum state of the object and the number of charges
transferred through the junction.  In the previous paper, the
non-Markovian master equation for the charge specific density matrix
for an arbitrary quantum object coupled to a low transparency tunnel
junction was derived.\cite{RamSheWab04} A non-Markovian master
equation is less tractable for calculational purposes and the
Markovian approximation is employed in the present paper.  This is
quite sufficient for calculations of average properties, such as the
average current through the tunnel junction, where only the long time
behavior needs to be addressed. However, when calculating the current
noise of the junction, the Markovian approximation limits the
description to the low frequency noise as discussed in section
\ref{corrr}.

The Markovian charge specific
master equation for a quantum object coupled to the junction
 was in general shown to have
the form\cite{RamSheWab04}
\begin{equation}
\dot{\hat{\rho}}_{n}(t)=
\frac{1}{i\hbar}[\hat{H}_{0},\hat{\rho}_{n}(t)]+\Lambda\{\hat{\rho}_{n}(t)\}
+\mathcal{D}\{\hat{\rho}_{n}''(t)\}
+\mathcal{J}\{\hat{\rho}_{n}'(t)\},
\label{Xconmeq}
\end{equation}
where $\hat{\rho}_{n}'$ and $\hat{\rho}_{n}''$ denote the ``discrete
derivatives'' introduced in Eqs.~(\ref{firstderiv}, \ref{deriv}),
and the Lindblad-like super operator, $\Lambda\{\hat{\rho}_{n}\}$,
has the form 
\begin{eqnarray}
\Lambda\{\hat{\rho}\}=\frac{1}{\hbar}\left[ \sum\limits
_{\mathbf{l}\mathbf{r}}f_{\mathbf{l}}(1-f_{\mathbf{r}})
\left(\left[\hat{T}_{\mathbf{l}\mathbf{r}}^{\dagger}\right]
\hat{\rho}\hat{T}_{\mathbf{l}\mathbf{r}}
-\hat{T}_{\mathbf{l}\mathbf{r}}\left[\hat{T}_{\mathbf{l}
\mathbf{r}}^{\dagger}\right]\hat{\rho}\right)\right.\\\label{lambda}
+\left.\sum_{\mathbf{l}\mathbf{r}}f_{\mathbf{r}}(1-f_{\mathbf{l}})
\left(\hat{T}_{\mathbf{l}\mathbf{r}}\hat{\rho}
\left[\hat{T}_{\mathbf{l}\mathbf{r}}^{\dagger}\right]
-\hat{\rho}\left[\hat{T}_{\mathbf{l}\mathbf{r}}^{\dagger}\right]
\hat{T}_{\mathbf{l}\mathbf{r}}\right)\right]+H.c.
\nonumber 
\end{eqnarray}
where here and in the following $H.c.$ represents the hermitian
conjugate term with respect to the variable of the quantum object.
The bracket denotes the operation
\begin{equation}
\left[\hat{T}_{\mathbf{lr}}\right]=\frac{1}{\hbar}\int\limits
_{0}^{\infty}d\tau
e^{i(\varepsilon+eV)\tau/\hbar}e^{-i\hat{H}_{0}\tau/\hbar}
\hat{T}_{\mathbf{lr}}e^{i\hat{H}_{0}\tau/\hbar},
\label{evol}
\end{equation}
where $\hat{T}_{\mathbf{lr}}$ is the oscillator perturbed tunneling
amplitude, Eq.~(\ref{tlin}), and
$\varepsilon=\varepsilon_{\mathbf{l}}-\varepsilon_{\mathbf{r}}$, and
$f_{\mathbf{l}}$ and $f_{\mathbf{r}}$ are the single particle energy
distribution functions for the electrodes which in the following are
assumed in equilibrium described by the junction temperature $T$.  In
this paper, we restrict ourself to the case where the junction is
biased by a constant voltage $U$, denoting $V=eU$, where $e$ is the
electron charge.  The dagger indicates hermitian conjugation of
operators of the coupled quantum object.

The drift super-operator is
\begin{eqnarray}
\mathcal{J}\{\hat{R}\}
&=&\frac{1}{\hbar}\sum\limits
_{\mathbf{l}\mathbf{r}}\frac{F_{\mathbf{l}\mathbf{r}}^{a}}{2}
\left(\left[\hat{T}_{\mathbf{l}\mathbf{r}}^{\dagger}\right]
\hat{R}\hat{T}_{\mathbf{l}\mathbf{r}}
+\hat{T}_{\mathbf{l}\mathbf{r}}\hat{R}
\left[\hat{T}_{\mathbf{l}\mathbf{r}}^{\dagger}\right]
+H.c\right)\nonumber \\
&+&\frac{F_{\mathbf{l}\mathbf{r}}^{s}}{2}\left(
\left[\hat{T}_{\mathbf{l}\mathbf{r}}^{\dagger}\right]
\hat{R}\hat{T}_{\mathbf{l}\mathbf{r}}
-\hat{T}_{\mathbf{l}\mathbf{r}}\hat{R}
\left[\hat{T}_{\mathbf{l}\mathbf{r}}^{\dagger}\right]+H.c.\right),
\label{Xopdrift}
\end{eqnarray}
and it has been written in terms of the symmetric and antisymmetric
combinations of the distribution functions 
\[
F_{\mathbf{l}\mathbf{r}}^{s}=f_{\mathbf{l}} 
+f_{\mathbf{r}}-2f_{\mathbf{l}}f_{\mathbf{r}},\qquad
F_{\mathbf{l}\mathbf{r}}^{a}=f_{\mathbf{l}}-f_{\mathbf{r}}.
\]

The diffusion super-operator can be obtained from the drift
super-operator according to
\begin{equation}
\mathcal{D}\{\hat{R}\}=\frac{1}{2}
\mathcal{J}_{s\leftrightarrow a}\{\hat{R}\},
\label{Xopdiff}
\end{equation}
where the subscript indicates that symmetric and antisymmetric
combinations of the distribution functions should be interchanged, $
F_{\mathbf{l}\mathbf{r}}^{s}\leftrightarrow
F_{\mathbf{l}\mathbf{r}}^{a}$.

The bracket notation is specified in Eq.~(\ref{evol}), where in
general $H_{0}$ denotes the Hamiltonian for the isolated arbitrary
quantum object. In the following we consider an oscillator coupled to
the junction, and $H_{0}$ represents the isolated harmonic oscillator.

As expected, the coupling of the oscillator to the tunnel junction
leads to a renormalization of its frequency,
$\omega_{B}^{2}\rightarrow\omega_{0}^{2}$.  The renormalization
originates technically in the term present in the Lindblad-like
operator, $\Lambda\{\hat{\rho}\}$, which is quadratic in the
oscillator coordinate and of commutator form with the charge specific
density matrix, and gives for the renormalized frequency
\[
\omega_{0}^{2}=\omega_{B}^{2}-
\frac{2}{m}\sum_{\mathbf{lr}}F_{\mathbf{l}\mathbf{r}}^{a}(P_{+}
+P_{-})|w_{\mathbf{lr}}|^{2}
\]
where
\begin{eqnarray}
P_{\pm}=\Re\frac{1}{\varepsilon-V\mp\hbar\omega_0 +i0}.
\label{princ05}
\end{eqnarray}

For the considered interaction, the renormalization can be simply
handled by changing in Eq.~(\ref{evol}) from the evolution by the bare
oscillator Hamiltonian to the oscillator Hamiltonian with the
renormalized frequency, the shift being compensated by subtracting an
identical counter term. The above frequency shift is then identified
by the counter term having to cancel the quadratic oscillator term of
commutator form generated by the $\Lambda\{\hat{\rho}\}$-part in
Eq.~(\ref{Xconmeq}). Substituting for the bare oscillator Hamiltonian,
$\hat{H}_{0}$, in (\ref{evol}) the renormalized Hamiltonian,
Eq.~(\ref{hren}), all quantities are expressed in terms of the
physically observed oscillator frequency $\omega_{0}$. In particular, the
bracket becomes
\begin{widetext}
\begin{equation}
 \left[\hat{T}_{\mathbf{l}\mathbf{r}}\right]
 =\pi\left(\delta_{0}v_{\mathbf{lr}}+(\delta_{+}
 +\delta_{-})\frac{w_{\mathbf{lr}}}{2}\hat{x}
 -i(\delta_{+}-\delta_{-})\frac{w_{\mathbf{lr}}}{2m\omega_0}\hat{p}\right)
  +i\left(P_{0}v_{\mathbf{lr}}+(P_{+}
 +P_{-})\frac{w_{\mathbf{lr}}}{2}\hat{x}
 -i(P_{+}-P_{-})\frac{w_{\mathbf{lr}}}{2m\omega_0}\hat{p}\right),
 \label{tbr}
 \end{equation}
\end{widetext}
where 
\begin{eqnarray}
\delta_{0}=\delta(\varepsilon-V),\qquad
\delta_{\pm}=\delta(\varepsilon-V\mp\hbar\omega_0),
\label{delt}
\end{eqnarray}
and
\begin{eqnarray}
P_{0}=\Re\frac{1}{\varepsilon-V+i0}.
\label{princ}
\end{eqnarray}
In the following the notation $\Omega=\hbar\omega_0$ for the
characteristic oscillator energy is introduced.

Evaluating the diffusion and drift operators for the case of position
coupling of the oscillator to the junction, Eq.~(\ref{tlin}), we
obtain for the drift super-operator
\begin{widetext}
\begin{equation}
 \mathcal{J}\{\hat{R}\}  = 
 V\mathcal{G}\{\hat{R}\}+\frac{\hbar}{ m\Omega}G_{xx}
 \Delta_{V}\Im(\hat{x}\hat{R}\hat{p})
 + G_{x}\frac{i\hbar}{2m\Omega}\Delta_{V}[\hat{p},\hat{R}]+
 g_{x} \frac{\hbar}{2m}\{\hat{p},\hat{R}\}
    + 
   g_{x}\frac{1}{2i}\left(V\coth\frac{V}{2T}+S_{V}\right)[\hat{x},\hat{R}] ,
 \label{2drift2}
 \end{equation}
\end{widetext}
where 
the conductance super-operator is defined
as
\begin{equation}
\mathcal{G}\{\hat{R}\}
\equiv G_{0}\hat{R}+ G_{x}\{\hat{x},\hat{R}\}+ G_{xx}\hat{x}\hat{R}\hat{x} ,
\label{w5d}
\end{equation}
and
\begin{equation}
\Delta_{V}=\frac{V+\Omega}{2}\coth\frac{V+\Omega}{2T}
-\frac{V-\Omega}{2}\coth\frac{V-\Omega}{2T},
\label{dv}
\end{equation}
and
\begin{equation}
S_{V}=\frac{V+\Omega}{2}\coth\frac{V+\Omega}{2T}
+
\frac{V-\Omega}{2}\coth\frac{V-\Omega}{2T},
\label{}
\end{equation}
the latter being proportional to the current noise power spectrum at
the frequency of the oscillator.  For the diffusion super-operator we
obtain
\begin{widetext}
\begin{eqnarray}
\mathcal{D}\{\hat{R}\}  &=& 
\frac{V}{2}\coth\frac{V}{2T}\mathcal{G}\{\hat{R}\} +
\frac{G_{xx}}{2}\left(B_{V}\hat{x}\hat{R}\hat{x}+
\frac{\hbar}{m}\Im(\hat{x}\hat{R}\hat{p})
\right)+\frac{\hbar A}{m\Omega}\Re(\hat{p}\hat{R}\hat{x})\nonumber \\
   && \quad 
  +\frac{G_{x}}{4}\left(B_{V}\{\hat{x},\hat{R}\}
+\frac{i\hbar}{m}[\hat{p},\hat{R}]\right)  
   +  g_{x}\left(\frac{V}{2i}[\hat{x},\hat{R}]
+\frac{\hbar \Delta_{V}}{4 m\Omega}\{\hat{p},\hat{R}\}\right),
\label{2diff2}
\end{eqnarray}
\end{widetext}
where
\begin{equation}
B_{V}= S_{V} - V\coth\frac{V}{2T}
\label{2sv}
\end{equation}
and we have introduced the notation
\begin{equation}
\Re(\hat{p}\hat{R}\hat{x})=\frac{1}{2}\left((\hat{p}\hat{R}\hat{x})+(\hat{p}\hat{R}\hat{x})^{\dagger}\right),
\end{equation}
and
\begin{equation}
\Im(\hat{p}\hat{R}\hat{x})=\frac{1}{2i}\left((\hat{p}\hat{R}\hat{x})-(\hat{p}\hat{R}\hat{x})^{\dagger}\right),
\end{equation}
in  (\ref{2drift2}) and (\ref{2diff2}).

The parameter $A$ in Eq.~(\ref{2diff2}) is given by
\begin{equation}
A_V=\frac{\hbar^{2}}{2m\Omega}
\sum_{\mathbf{lr}}|w_{\mathbf{lr}}|^{2}
F_{\mathbf{l}\mathbf{r}}^{s}(P_{+}-P_{-}),
\label{ain}
\end{equation}
and was encountered and discussed in connection with the unconditional
master equation, Eq.~\ref{meqfin}.  Technically, it originates in our
model from the principal value of integrals, i.e., from virtual
processes where electronic states far from the Fermi surface are
involved.  Estimating its magnitude under the assumption that the
couplings $|w_{\mathbf{lr}}|^2$ are constants, one obtains with
logarithmic accuracy
\begin{equation}
A \approx \frac{2\hbar^2  G_{xx}}{\pi m}
\ln \frac{E_F}{\rm{max}(V,T,\Omega)}
\; .
\label{afin}
\end{equation}

In the course of evaluating the diffusion and drift operators,
combinations like
$\Im/\Re\left(v_{\mathbf{lr}}^{*}w_{\mathbf{lr}}\right)$ appear
together with principal value terms. The phase of
$v_{\mathbf{lr}}^{*}w_{\mathbf{lr}}$ will in general be a random
function of the electron reservoir quantum numbers $\mathbf{l}$ and
$\mathbf{r}$.  Summing over these quantum numbers, where the principal
value term does not provide any restriction of the energy interval, as
it happens in the case of terms proportional to delta functions, they
will tend to average to zero, and we shall therefore in the following
neglect such terms.\cite{fterm}

\section{ Noise in Markovian approximation}\label{tech}

In this section we evaluate in the Markovian approximation the
current-current correlator of the tunnel junction for the case of a
harmonic oscillator coupled to the junction.  The task has been
reduced to evaluating the expressions in Eq.~(\ref{qc2}) and
Eq.~(\ref{rc2}), i.e., quantities of the form Eq.~(\ref{sc2}) where
the involved super-operator is the evolution operator for the charge
unconditional density matrix given in Eq.~(\ref{meqfinfin}).

It immediately follows from the master equation,
Eq.~(\ref{meqfinfin}), that quantities like $X(t)= {\rm Tr }
(\hat{X}\hat{\rho}(t))$, where $ \hat{X}$ 
can denote $\hat{x}$, $\hat{p}$, $\hat{x}^2$, and
$\hat{p}^{2}$, and $\hat{\rho}$ is an arbitrarily normalized solution to the
master equation, satisfy the corresponding classical equations of
motion for a damped oscillator.  The variables entering
Eq.~(\ref{sc2}) can therefore be expressed in terms of their initial
values at time $t = 0$, and restricting ourselves for simplicity to
the case of weak damping, $\gamma \ll \omega_{0}$, they have the form
corresponding to that of an underdamped classical oscillator
\begin{equation}
x_{J}(t)= x_{J}(0) e^{- \gamma t}
\cos \omega_{0} t 
+
\frac{p_{J}(0)}{m\omega_{0} }e^{- \gamma t}\sin \omega_{0} t
\label{tc2}
\end{equation}
and
\begin{equation}
 p_{J}(t)=
 -m\omega_{0}  x_{J}(0) e^{- \gamma t} \sin \omega_{0}t
 +
 p_{J}(0)e^{- \gamma t}\cos \omega_{0} t
 \label{sae}
 \end{equation}
and
\begin{eqnarray}
x_{J}^2(t)
&=&  e^{- 2\gamma t} x_{J}^2(0)
\cos^2\omega_{0} t
+
\frac{p_{J}^2(0)}{m^2\omega_{0}^2}e^{- 2\gamma t}\sin^2\omega_{0} t \nonumber \\
      &+&   
 e^{- 2\gamma t}
\frac{ \{x,p\}_{J}(0)}{2m\omega_{0}}
\sin2\omega_{0} t .
\label{uc2}
\end{eqnarray}
The initial values in these equations are found from Eq.~(\ref{wc2})
to be
\begin{equation}
x_{J}(0) =
G_{x}\left(\frac{\hbar^2}{m \Omega } \right)
\left(
V  (2N^{*}+1)-
\frac{1}{2}
\Delta_{V}   \right)
\label{ya22}
\end{equation} 
\begin{equation}
\frac{p_{J}(0)}{m}  =
g_{x}\frac{\hbar}{2}\left(\Omega (2N^{*} + 1)-
\left(V\coth\frac{V}{2T}+S_{V}\right)\right)
\label{za22}
\end{equation}
\begin{eqnarray}
&& \frac{p_{J}^2(0)}{2m} 
+
\frac{m \omega_0^2 x_{J}^2(0)}{2}=\nonumber \\ 
&&\tilde{G}_{xx} 
\left(
\Omega V N^{*2} +
\Omega \left(V - \Delta_{V} \right) (N^{*}+ \frac{1}{2})
 \right)
\label{6c2}
\end{eqnarray}
\begin{equation}
\frac{m \omega_0^2 x_{J}^2(0)}{2} 
-
\frac{p_{J}^2(0)}{2m}
=
\tilde{G}_{xx}
\;
\Omega V N^{*} (N^{*}+1)
\label{7c2}
\end{equation}
\begin{equation}
\{x,p\}_{J}(0) =0
\; .
\label{tae}
\end{equation}
Substituting these initial values into equations
 Eqs.~(\ref{tc2}-\ref{uc2}), we obtain using Eqs.~(\ref{qc2}) and
 (\ref{rc2}) the expressions in Eq.~(\ref{ac2}) for the regular part of the
 current-current correlator
\begin{widetext} 
\begin{eqnarray}
 S_{x}(t)      &=& 
 \tilde{G}_{x}^2
 e^{-\gamma t}\cos \omega_{0} t
 V
 \left(
 2V(2N^{*} +1) -
 \Delta_{V}   \right)-  
  \tilde{g}_{x}^2
 e^{-\gamma t}\cos \omega_{0} t 
  \left(
 \frac{1 }{2}\Omega V\coth\frac{V}{2T}+ \Omega^2(N_{e}-N^{*})\right)\nonumber \\
       &&-   
 \tilde{g}_{x}\tilde{G}_{x}
 e^{-\gamma t}\sin \omega_{0} t
  \left(
 V^2\coth\frac{V}{2T} + \Omega V (2N_{e}+1) - \frac{1}{2}\Omega \Delta_{V}
  \right)
\label{bla23}
\end{eqnarray}
 and
\begin{equation}
  S_{x^2}(t) =   
 \frac{1}{2}
 \tilde{G}_{xx}^2
 e^{-2\gamma t}
  V\left(
 2V N^{*2} +
  \left(V - \Delta_{V} \right) (2N^{*}+1)
  \right)
        +  
 \tilde{G}_{xx}^2
 e^{-2\gamma t}
 V^2N^{*}(N^{*}+1)\cos 2\omega_{0}t
 \; .
 \label{7b23}
 \end{equation}
\end{widetext}
Evaluating in Eq.~(\ref{q9d}) the trace of the diffusion
super-operator in the stationary state of the oscillator, ${\cal
D}\left\{ \hat{\rho}_{s}\right\}$, we obtain according to
Eq.~(\ref{2diff2}) for the singular part of the noise correlator
\begin{widetext}
\begin{equation}
S_{1}(t) =
2\delta(t)
\left( 
G_{0}
\frac{V}{2}\coth\frac{V}{2T} 
+\tilde{G}_{xx}
\frac{\Omega }{2}
\left(
N^{*}(N_{e}+1) + N_{e}(N^{*}+1)
\rule[-1ex]{0ex}{0ex}
\right)
\right)
\; .
\label{pc2} 
\end{equation}
\end{widetext}
We observe that as to be expected, the Markovian approximation for the
dynamics of the charge specific density matrix only captures the
low-frequency noise, $\omega~<~{\rm max}\left(\frac{T}{\hbar },
\frac{V}{\hbar }\right)$, and in section \ref{corrr} we shall
therefore only discuss this limit.\cite{nonmark}


\begin{thebibliography}{100}
\bibitem{virial}It is interesting to note that the virial theorem
turns out to be violated due to the
coupling. Indeed, the renormalization in Eq.~(\ref{ja2}) modifies
the potential energy, $V_{\textrm{pot}}=\frac{m\omega^{2}x^{2}}{2}$,
but does not influence the kinetic energy $K=\frac{p^{2}}{2m}$. 
As a result, $\langle V\rangle-\langle K\rangle=\frac{1}{2}A$, in
disaccord with the virial theorem for a harmonic oscillator,
$\langle V\rangle = \langle K\rangle$ .
In fact, this is an exact result in the sense that it can be obtained,
without reference to the renormalization procedure, directly from
Eq.~(\ref{meqfin}) by the standard Ehrenfest construction of the
equation of motion for expectation values.
\bibitem{MacD}The current-current correlator, $S(t)$, tends to zero at
  large times and a convergence factor, $\exp(-\delta t)$,
can be inserted into the integrand in the expression for the power spectrum,
Eq.~(\ref{kd2}). Upon inserting
the expression for the current-current correlator, Eq.~(\ref{i8d}), 
into the expression for the
power spectrum, a partial integration gives 
MacDonald's formula
$
S_{\omega}= 2\omega \int_{0}^{\infty}dt \,e^{-\delta t} \sin(\omega t)
\frac{d\left\langle \left\langle n^{2}(t)\right\rangle \right\rangle
}{dt}
$,
since the vanishing of the boundary term at $t=0$
is ensured by virtue of Eq.~(\ref{j8d}).
\bibitem{fterm} A similar renormalization
was encountered and discussed in connection with a two-level system
coupled to a tunnel junction.\cite{RamSheWab04}
\bibitem{nonmark}
We recall the result for an isolated junction:\cite{RamSheWab04}
using the non-Markovian master equation for the charge
specific density matrix,
one obtains for the power spectrum of current noise 
$
S_{\omega} =
I(V+ \omega ) \coth \frac{V+ \omega }{2T}
+
I(V - \omega ) \coth \frac{V- \omega }{2T}
$,
the general result for a biased junction, valid
for arbitrary frequency and non-linear $I-V$ characteristic, $I=
I(V)$. The Markovian approximation only captures the low frequency
limit of the noise power.
\end{thebibliography}
\end{document}